\documentclass[aps,prd,twocolumn,superscriptaddress,longbibliography,nofootinbib]{revtex4-2}
\usepackage[utf8]{inputenc}
\usepackage{xcolor}
\usepackage{graphicx}
\usepackage{subfigure}
\usepackage{multirow}
\usepackage{tensor}

\usepackage{amsmath,amssymb,amsfonts}

\begin{document}

%=========================================================
\title{Evolution of Brill waves with an adaptive pseudospectral method}
%=========================================================

\author{Isabel Su\'arez Fern\'andez}
\affiliation{
  Centro de Astrof\'{\i}sica e Gravita\c c\~ao -- CENTRA,
  Departamento de F\'{\i}sica, Instituto Superior T\'ecnico -- IST,
  Universidade de Lisboa -- UL, Av.\ Rovisco Pais 1, 1049-001 Lisboa,
  Portugal}

\author{Sarah Renkhoff}
\affiliation{Friedrich-Schiller-Universität,
Jena, 07743 Jena, Germany}

\author{Daniela Cors Agulló}
\affiliation{Friedrich-Schiller-Universität,
Jena, 07743 Jena, Germany}

\author{Bernd Brügmann}
\affiliation{Friedrich-Schiller-Universität,
Jena, 07743 Jena, Germany}

\author{David Hilditch}
\affiliation{
  Centro de Astrof\'{\i}sica e Gravita\c c\~ao -- CENTRA,
  Departamento de F\'{\i}sica, Instituto Superior T\'ecnico -- IST,
  Universidade de Lisboa -- UL, Av.\ Rovisco Pais 1, 1049-001 Lisboa,
  Portugal}

\begin{abstract}
As a first application of the new adaptive mesh functionality of the
the pseudospectral numerical relativity code \texttt{bamps}, we evolve
twist-free, axisymmetric gravitational waves close to the threshold of
collapse. We consider six different one-parameter families of Brill
wave initial data; two centered and four off-centered families. Of
these the latter have not been treated before. Within each family we
tune the parameter towards the threshold of black hole formation. The
results for centered data agree with earlier work. Our key results
are first, that close to the threshold of collapse the global peak in
the curvature appears on the symmetry axis but away from the origin,
indicating that in the limit they will collapse around disjoint
centers. This is confirmed in three of the six families by explicitly
finding apparent horizons around these large curvature peaks. Second,
we find evidence neither for strict discrete-self-similarity nor for
universal power-law scaling of curvature quantities. Finally, as in
Ledvinka \& Khirnov's recent study, we find approximately universal
strong curvature features. These features appear multiple times within
individual spacetimes close to the threshold and are furthermore
present within all six families.
\end{abstract}

\maketitle

%=========================================================
\section{Introduction}
\label{sec:intro}
%=========================================================

Working in spherical symmetry with a massless scalar field minimally
coupled to the Einstein field equations, and tuning in solution space
to the verge of black hole formation, Choptuik~\cite{Cho93} observed
behavior with a striking resemblance to that observed near critical
points in other areas of physics. For instance, near a thermodynamic
critical point, as the correlation length diverges the system is
rendered scale invariant. Order parameters then follow power-laws with
universal critical exponents. Universality, power-law behavior, and
scale invariance (also called self-similarity) are together referred
to as critical phenomena. The discovery of all three in gravitational
collapse are thus known as critical phenomena in gravitational
collapse.

Despite the familiarity of the phenomena, what is meant as the
critical point in gravitational collapse has its own particular
definition. The threshold of collapse can be conveniently identified
by considering families of initial data labeled by a
single-parameter~$p$. The parameter varying within a family is
conventionally chosen such that for small values of the parameter ($p
< p_{\star}$), the field content will eventually disperse and lead
asymptotically to flat space. On the other hand large values of the
parameter ($p > p_{\star}$), lead to black hole formation and thus the
presence of a singularity of some description. Of deep interest is the
threshold of collapse~$p=p_{\star}$. Within the analogy to statistical
physics, the critical point is precisely the threshold of collapse.

The manifestation of critical phenomena in gravitational collapse have
their own specifics. Regarding for instance scale-invariance, by using
a suitable time coordinate close to the threshold of collapse,
Choptuik observed progressively scaled down, periodic repetitions of a
field configuration that appeared independently of the family under
consideration. Closer to the threshold more and more of these {\it
  echoes} were found. This lead to the conjecture that around the
center of collapse, the threshold spacetime within any family agrees
with that of any other, and that this limiting spacetime itself is
discretely self-similar (DSS). The limiting spacetime is therefore
referred to as the critical solution or Choptuik spacetime. Such a
spacetime has been proven to exist~\cite{ReiTru19}. Universality was
also observed in the periodicity of these echoes, as well as in the
exponent, $\gamma$, of the power-law that the masses of near critical
black holes obey,
\begin{align}\label{eq:m_bh_power_law}
	M_{\text{BH}}\simeq |p-p_{\star}|^{\gamma}.
\end{align}
In~\cite{GarDun98}, it was argued on dimensional grounds and confirmed
empirically that below the threshold an analogous scaling relation
should hold for the spacetime maximum of curvature scalars such as the
Kretschmann scalar~$I = R_{abcd}{}R^{abcd}$, so that
\begin{align}\label{eq:I_power_law}
	I_{\text{max}}{}^{-1/4}\simeq|p - p_{\star}|^{\gamma}.
\end{align}
Arguments of~\cite{Gun97,HodPir97} suggest that if the critical
spacetime is DSS then this power-law should occur with a periodic
wiggle superimposed. The above properties have been repeatedly
verified for the spherical massless scalar field for many different
families of data. In spherical symmetry various alternative matter
models have been considered and, although details such as the values
of the power-law parameters or the specific type of self-similarity
differ, the basic ingredients persist (see the canonical
review~\cite{GunGar07} for details).

Less clear is the precise extent to which critical phenomena extend
beyond spherical symmetry. In line with the conjecture that there is a
single strong-field solution lying between dispersion and collapse,
for the original matter model of a massless scalar field it is
known~\cite{GarGun98} that nonspherical linear mode perturbations
about the critical solution decay. On the other hand nonlinear
numerical work in axisymmetry~\cite{ChoHirLie03,Bau18}, which is much
more challenging than the spherical setting, indicates that both
power-law behavior and periodicity vary in the presence of increasing
aspherical perturbations, and furthermore that the single center of
collapse generically splits into two distinct centers resembling to an
extent the Choptuik spacetime. There are even a small number of
studies in full-3d \cite{HeaLag13, DepKidSch18}, which are again
correspondingly more numerically challenging and thus remain further
from the threshold. Keeping the axisymmetric setup but considering
electromagnetic waves~\cite{BauGunHil19,MenBau21} as a matter model
has proven to show the exact same difficulty in establishing strict
DSS and universal critical exponents, and again the appearance of
multiple centers of collapse. Evolutions of complex scalar field
content with angular momentum also suggests a more complicated
structure in phase space~\cite{ChoHirLie03}. Toy
models~\cite{SuaVicHil21} also suggest that universality could be more
subtle as one moves away from spherical symmetry and more parameters
are needed to parameterize threshold solutions. These issues demand
deeper investigation in axisymmetry and full-3d.

As the simplest scenario in which dynamical gravitational waves occur,
the aforementioned axisymmetric setting is of special interest in
general relativity (GR). By working in vacuum we can try to understand
the aspect of critical collapse determined by pure gravity. Two main
types of vacuum initial data have been considered: Brill and Teukolsky
waves. The former were introduced by Brill~\cite{Bri59} and refer to a
solution of the constraint equations in axisymmetric vacuum at a
moment of time symmetry. The latter, constructed first by
Teukolsky~\cite{Teu82}, and generalized from quadrupolar to include
all multipoles by Rinne~\cite{Rin08b}, are general vacuum solutions to
the linearized Einstein field equations in transverse-traceless gauge.
For a comparison between these two types of waves at the linear level
see~\cite{SuaBauHil21}. As linearized solutions, Teukolsky wave
initial data has to be dressed up to construct solutions of the
constraint equations of GR. There are several strategies to do
so~\cite{AbrEva93,AbrEva94,ShiNak95,BonGouGra03,PfeKidSch04,HilBauWey13}.

Concerning vacuum time evolutions, Abrahams and
Evans~\cite{AbrEva93,AbrEva94} were the first to study gravitational
waves near the threshold of collapse already in the early
1990s. Following the basic approach of Choptuik they evolved members
of one-parameter families of (constraint solved) Teukolsky waves
without moment of time symmetry and tuned to the threshold. They
observed echoes and found that for supercritical data the black hole
masses also obey the power-law~\eqref{eq:m_bh_power_law}. Since then,
numerous
authors,~\cite{AlcAllBru99a,GarDun00,San06,Rin08,Sor10,HilBauWey13,
  HilWeyBru17,KhiLed18,LedKhi21} have evolved gravitational waves to
investigate the threshold of collapse, but none have unambiguously
recovered this early success. Two of these deserve special attention
here. First, in~\cite{HilWeyBru17}, the direct precursor to the
present work, a single family of Brill waves was considered. In
qualitative agreement with other axisymmetric
simulations~\cite{ChoHirLie03,Bau18,BauGunHil19,MenBau21}, close to
the threshold global maxima of the curvature form away from the origin
as the consequence of large spikes appearing in the curvature. In
collapse spacetimes, a disjoint pair of horizons were found along the
symmetry axis around the largest of the foregoing curvature
spikes. Tentative evidence for power-law scaling with a periodic
wiggle was found for the Kretschmann scalar, but nothing could be
concluded about universality with only one family. In important recent
work, Ledvinka and Khirnov~\cite{LedKhi21} directly confirmed these
findings using a different gauge and completely independent code, and
tuning to a comparable neighborhood of the threshold. Studying now
multiple families of initial data, they furthermore found pairs of
apparent horizons (AHs) also in (constraint solved) Teukolsky
waves. Once more in qualitative agreement with other work in
axisymmetry, they also report that although power-law scaling does
appear within given families, the {\it power} itself does not seem to
be universal. And there furthermore seems to be no universal threshold
solution. That said, the most important finding of~\cite{LedKhi21} is
the presence of repeated curvature features which although not
appearing in a strictly DSS fashion, {\it do} appear to be universal.

The key obstacle encountered by both~\cite{HilWeyBru17}
and~\cite{LedKhi21} was in tuning~$p$ to~$p_\star$. For context, in
spherical symmetry, either by using adaptive mesh refinement (AMR) as
in~\cite{Cho93} or well-chosen coordinates as in~\cite{GarDun98}, it
is often possible to tune to machine precision~$|p-p_\star|\sim
10^{-15}$. In axisymmetric matter evolutions, values
like~$|p-p_\star|\sim 10^{-9}$ are often possible, whereas the best
that could be managed in~\cite{HilWeyBru17} was~$|p-p_\star|\sim
10^{-5}$. There are several reasons for this. With less symmetry the
numerical cost to reach a given level of accuracy is necessarily
higher. Despite using a pseudospectral numerical method adapted to
axisymmetry~\cite{Bru11,HilWeyBru15}, the computational cost of the
evolutions skyrocketed near the threshold of collapse as finer
curvature features form and more resolution is needed. For example
in~\cite{HilWeyBru17} approximately~$10^6$ core hours were used for a
single family. Besides that, it was found that near the threshold,
spacetimes simply could not be classified as the evolution would fail
before an AH could be found. It was unclear whether this was caused by
a lack of resolution, constraint violation rendering the solution
unphysical, or the formation of coordinate singularities.

Presently we return to the problem of vacuum critical collapse with
our pseudospectral code~\texttt{bamps}. To mitigate against
computational cost the code has undergone a major redesign
since~\cite{HilWeyBru17}, and now employs AMR. Here we present
evolutions of six different families of Brill wave initial data. As
was done previously, we evolved prolate and oblate centered Brill
waves. We also evolved four off-center families of Brill waves that
have not been treated before. The paper is structured as follows. In
section~\ref{sec:bamps} we give a brief overview of our formulation
and numerical methods. Complete details of our approach to AMR will be
given in a forthcoming sister paper. In section~\ref{sec:phys_mod} we
explain our initial data, approach to tuning and our new apparent
horizon finder. Our numerics are presented in
section~\ref{sec:results}, before a closing discussion in
section~\ref{sec:summary}.

%=========================================================
\section{Formulation and numerical setup}
\label{sec:bamps}
%=========================================================

The time evolution of the different sets of initial data were
performed using the~\texttt{bamps} code. To evolve the spacetime, we
use a first order reduction of the generalized harmonic gauge (GHG)
formulation~\cite{LinSchKid05} of the Einstein equations, with gauge
choices as described in~\cite{HilWeyBru15}. The field equations are
\begin{align}
  \partial_t \tensor{g}{_a_b} &=
  \beta^i \partial_i \tensor{g}{_a_b}
  - \alpha \tensor{\Pi}{_a_b}
  + \gamma_1 \beta^i \tensor{C}{_i_a_b} \nonumber\\
  \partial_t \tensor{\Phi}{_i_a_b} &=
  \beta^j \partial_j \tensor{\Phi}{_i_a_b}
  - \alpha \partial_i \tensor{\Pi}{_a_b}
  + \gamma_2 \alpha \tensor{C}{_i_a_b} \nonumber\\
  &\quad+ \frac{1}{2} \alpha n^c n^d
  \tensor{\Phi}{_i_c_d} \tensor{\Pi}{_a_b}
  + \alpha\gamma^{jk} n^c \tensor{\Phi}{_i_j_c}
  \tensor{\Phi}{_k_a_b} \nonumber\\
  \partial_t \tensor{\Pi}{_a_b} &=
  \beta^i \partial_i \tensor{\Pi}{_a_b}
  - \alpha \gamma^{ij} \partial_i \tensor{\Phi}{_j_a_b}
  + \gamma_1 \gamma_2 \beta^i \tensor{C}{_i_a_b} \nonumber\\
  &\quad+ 2 \alpha g^{cd}
  \left( \gamma^{ij} \tensor{\Phi}{_i_c_a} \tensor{\Phi}{_j_d_b}
  - \tensor{\Pi}{_c_a} \tensor{\Pi}{_d_b}
  - g^{ef} \tensor{\Gamma}{_a_c_e} \tensor{\Gamma}{_b_d_f} \right)
  \nonumber\\
  &\quad - 2 \alpha \left( \tensor{\nabla}{_(_a} \tensor{H}{_b_)}
  + \gamma_4 \tensor{\Gamma}{^c_a_b} \tensor{C}{_c}
  - \frac{1}{2} \gamma_5 \tensor{g}{_a_b} \Gamma^c \tensor{C}{_c} \right)
  \nonumber\\
  &\quad-\frac{1}{2} \alpha n^c n^d \tensor{\Pi}{_c_d} \tensor{\Pi}{_a_b}
  - \alpha n^c \gamma^{ij} \tensor{\Pi}{_c_i} \tensor{\Phi}{_j_a_b}
  \nonumber\\
  &\quad+\alpha \gamma_0 \left( 2 \tensor{\delta}{^c_{(a}} \tensor{n}{_{b)}}
  - \tensor{g}{_a_b} n^c \right) \tensor{C}{_c}\,.
\end{align}
Here we denote spacetime component indices by Latin letters starting
from~$a$, with spatial components starting from~$i$, and use the
standard notation for the lapse, shift, spatial metric and spacetime
metric. The reduction constraint associated with the first order
reduction is given by~$C_{iab}=\partial_ig_{ab}-\Phi_{iab}$. The
harmonic constraint~$C_a$ is easily expressed as a combination of our
evolved variables~$g_{ab},\Phi_{iab,}\Pi_{ab},$
(see~\cite{LinSchKid05}). The constraint damping parameters were taken
to be~$\alpha\gamma_0=-\gamma_1=\gamma_2=2\gamma_4=2\gamma_5=1$
throughout. The GHG formulation comes with the freedom to choose gauge
source functions~$H^a$. We choose
\begin{align}
 n_aH^a &= -\eta_L \log(\gamma^{p/2}/\alpha)\,,\nonumber\\
 \gamma^i{}_aH^a &= - \eta_S/\alpha^2 \beta^i\,,\label{eqn:Gauge_Sources}
\end{align}
with free scalar functions~$\eta_L, \eta_S$. The free parameters were
initially lifted from the best results of~\cite{HilWeyBru17} and
adjusted as seemed appropriate from there. As such we started
with~$p=1,\eta_L=\bar{\eta}_L/\alpha^2=0.4/\alpha^2$ and~$\eta_S=6$. At
the outer boundary we employ radiation controlling, constraint
preserving boundary conditions described in~\cite{Rin06a}, with our
own adjustments described in~\cite{HilWeyBru15}.

Brill wave initial data (see section~\ref{subsec:brill}) was
constructed externally and then interpolated onto the computational
domain. \texttt{bamps} is a full-3d numerical relativity code, and
development tests are performed without symmetry. In the treatment of
axisymmetric data such as Brill waves however, we suppress one spatial
dimension using the cartoon-method~\cite{AlcBraBru99a}, with our
implementation following rather the approach of~\cite{Pre05}, so that
only two-dimensional data on the $x$-$z$-plane is evolved. We further
only consider the region where~$x > 0$ and~$z > 0$, as the other
regions are then given by the symmetry of the problem.

The time evolution is performed with the method of lines, using a
standard RK4 timestepping algorithm. Within \texttt{bamps}, the
computational domain is divided into patches in a \emph{cubed-ball}
fashion, forming a ball comprised of deformed cubes, each being
represented internally as a unit cube~$(u, v, w)^T \in [-1,1]^3$ in
the patch-local coordinates.

These patches are further divided into smaller grids, which contained
between~$21$ to~$31$ points in each dimension in the present
work. Within each grid, the numerical solution is represented by a
nodal spectral representation in a Chebyshev basis, using
Gauss-Lobatto collocation points in each dimension. The grids are
coupled to each other using a penalty method as described
in~\cite{Hes00,HesGotGot07,TayKidTeu10}.

\begin{figure}[t]
\includegraphics[scale=0.25]{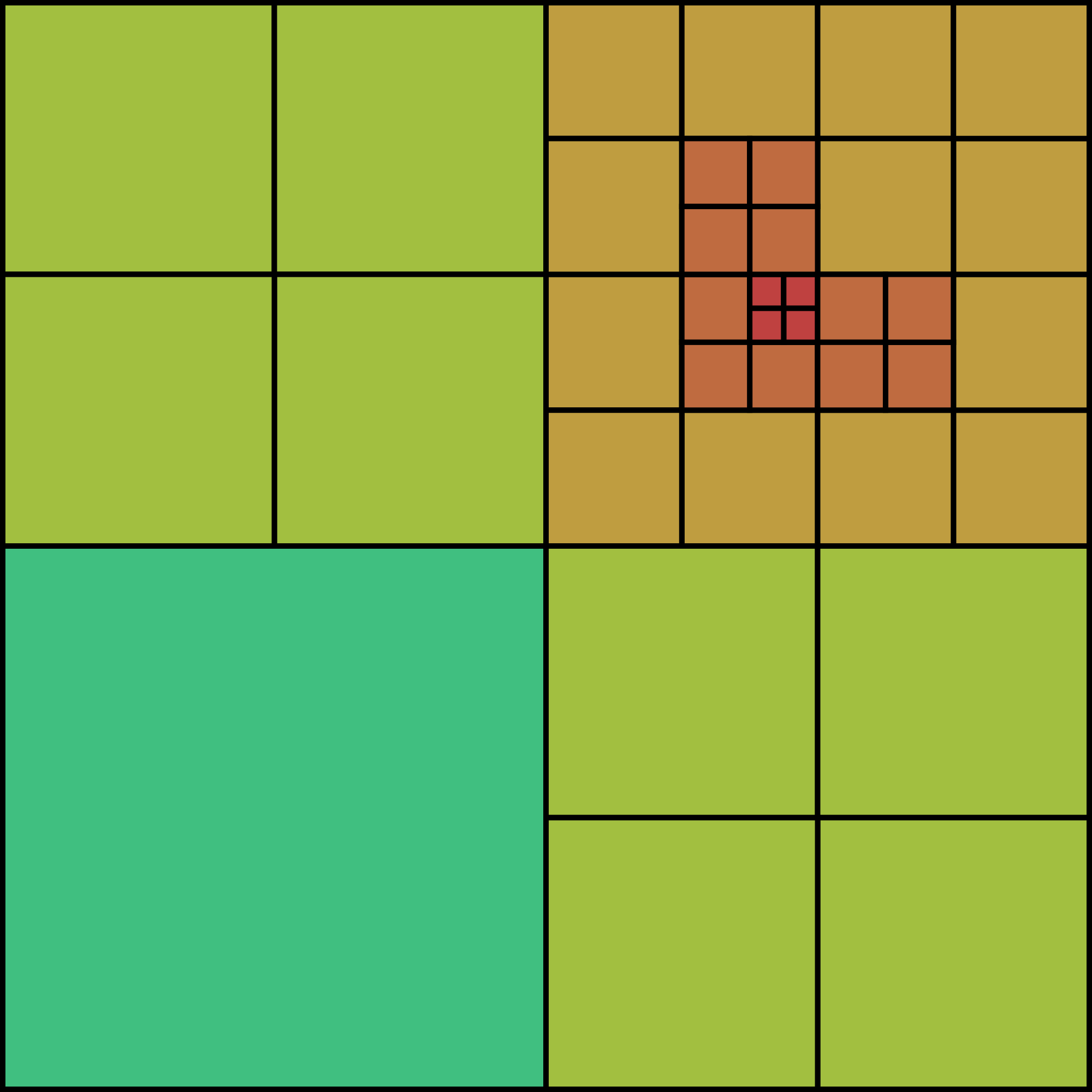}
\caption{Schematic example of a recursively refined grid structure, of
  the type \texttt{bamps} employs obeying a~$1:2$ condition. This
  example has~$5$ levels of refinement.\label{fig:amr-grids}
}
\end{figure}

The division of the patches into individual grids is driven by an AMR
system, which divides and combines grids based on heuristics that
estimate the quality of the data representation. This~$h$-refinement
is implemented subject to a~$1:2$ condition as illustrated in
Fig.~\ref{fig:amr-grids}. $p$-refinement functionality, in which the
number of points per individual grid is increased is also implemented,
but was not utilized in the work presented here. For most of our
simulations, an estimate of the truncation error of the spectral
series corresponding to the nodal data representation was used as the
refinement indicator, ensuring a grid-local relative truncation error
on the order of no more than~$10^{-9}$ throughout the domain, at least
up to the maximum number of refinement levels permitted ($19$ in this
work). A thorough technical description of our AMR setup is in
preparation.

All computations are parallelized at the grid-level, using MPI to
distribute computational work across processes. An integrated load
balancing system is employed to ensure an even distribution of both
computations and memory usage across all processes, redistributing
grids as needed whenever AMR operations cause a change in the grid
structure.

%=========================================================
%=========================================================

%=========================================================
\section{The Physical model} 
\label{sec:phys_mod}
%=========================================================

In this section we define the type of initial data we decided to
investigate. We then describe the process that allows us to estimate the
threshold amplitude within each of the evolved families. Finally we
include a description of the post-processing tools that have enabled
this estimation.

%=========================================================
\subsection{Brill waves as initial data}
\label{subsec:brill}
%=========================================================

Brill wave initial data are a gravitational-wave solution to vacuum
Einstein constraint equations. Following the procedure of
Brill~\cite{Bri59} it is possible to construct axisymmetric non-linear
initial data, where the spatial metric takes the form,
\begin{align}
  dl^2  = \gamma_{ij} dx^i dx^j = \Psi^4
  \left[ e^{2q} \left(d\rho^2 + dz^2 \right) + \rho^2 d\phi^2 \right]\,,
\label{eq:Brill_spatial_metric}
\end{align}
and, as the data are taken to be at a moment of time symmetry, the
extrinsic curvature vanishes identically. For the seed function we
always choose a general Gaussian
\begin{align}
  q(\rho,z) = A \rho^2 e^{- [(\rho - \rho_0)^2/\sigma_\rho^2 + (z-z_0)^2/\sigma_z^2]}
  \,,\label{eq:seed_function}
\end{align}
whose parameters differ depending on the studied family. In the
present work we will focus on six different families always with~$z_0
= 0$, $\sigma_\rho = \sigma_z = 1$ covering centered ($\rho_0 = 0$)
and off-centered ($\rho_0 = 4$ and $\rho_0 = 5$)
Gaussians both prolate ($A>0$) and
oblate ($A < 0$).

Of these six families, the first ($A > 0$, $\rho_0 = 0$) is best
studied in the literature, see for
example~\cite{HolMilWak93,AlcAllBru99a,GarDun00,HilWeyBru17,LedKhi21}.
In~\cite{LedKhi21} the centered oblate family was also treated, and so
these two families serve as a point of reference. Khirnov and Ledvinka
conclude that these centered families are more difficult to treat
numerically than their off-set (constraint solved) Teukolsky wave
initial data, which motivates our use of off-center Brill wave
data. (See~\cite{Rin08} for constrained evolution of alternative Brill
wave initial data, which we would also like to compare with in the
future).

%=========================================================
\subsection{Parameter search}
\label{subsec:search}
%=========================================================

%=========================================================
\begin{figure}[t]
\includegraphics[scale=0.56]{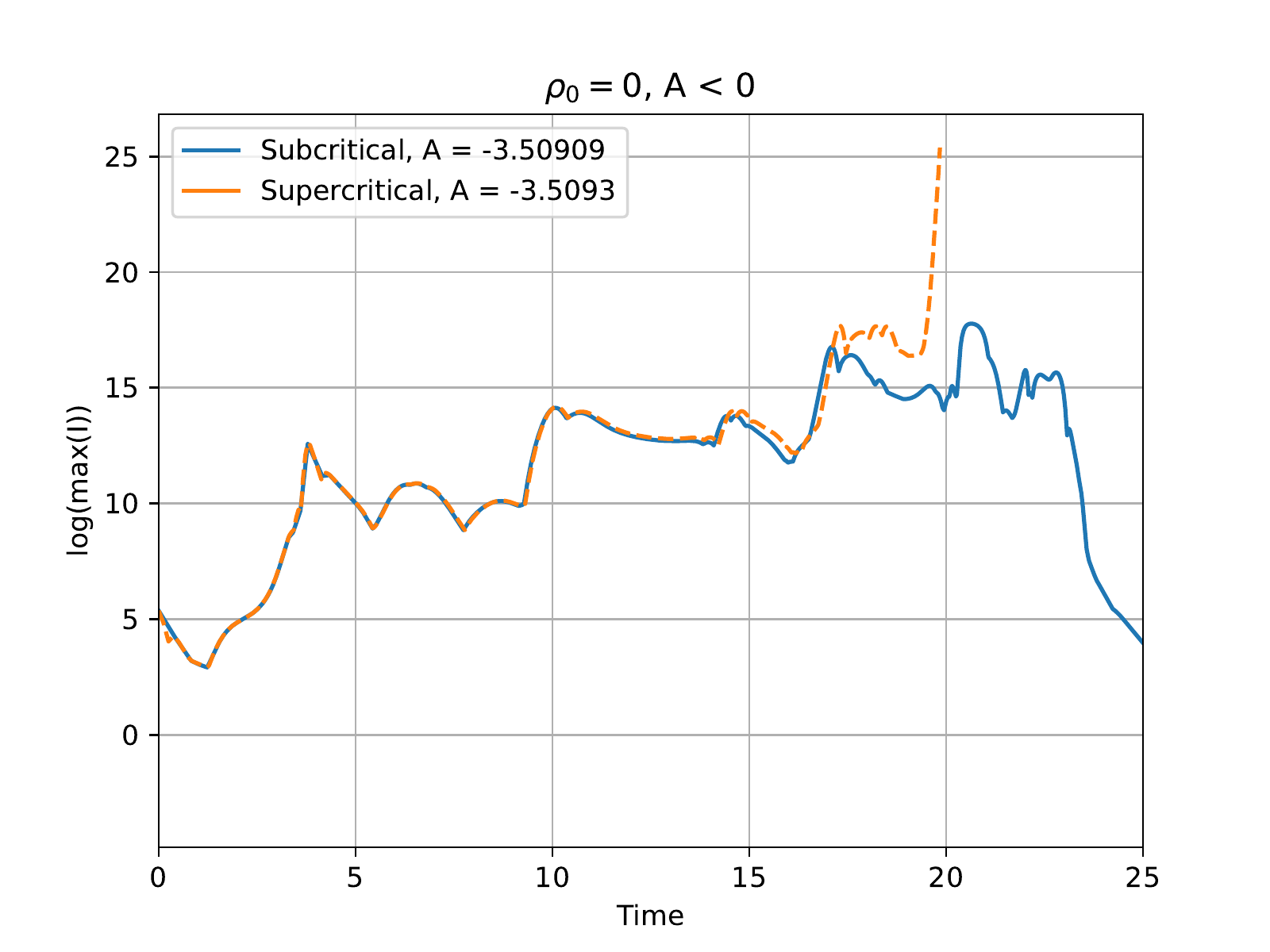}
\caption{An illustrative figure to show the different behavior for two
  close sets of initial data. The chosen family is a centered oblate
  Brill wave~$\rho_0 = 0$ with~$A < 0$. One can see that in the case
  of the orange dashed line, corresponding to a supercritical
  amplitude~$A = -3.50930$, the maximum of the Kretschmann scalar
  diverges. On the other hand, in the case of the subcritical
  amplitude~$A = -3.50909$ the Kretschmann's maximum decreases as the
  wave disperses.\label{fig:illustrative}}
\end{figure}
%=========================================================

In order to study phenomena near the threshold of collapse, the first
step is to identify that threshold as precisely as we can. With
one-parameter families, this is equivalent to identifying the value of
the parameter at the threshold. In our case, the parameter varying
within a family is the amplitude,~$A$, in
equation~\eqref{eq:seed_function}. Varying~$A$ allows us to classify
initial data depending on the outcome of its evolution. Initial data
whose evolution yields a dispersal of the fields is classified
as subcritical, and data that leads to black hole formation as
supercritical,
\begin{align}\label{eq:A_*}
	A_{\text{ subcritical}} < A_{\star} < A_{\text{ supercritical}}
\end{align}
(and similarly for families with~$A<0$).

In Fig.~\ref{fig:illustrative} one can see how the evolution of the
logarithm of the maximum of the Kretschmann scalar compares between a
subcritical and a supercritical run. This plot illustrates how similar
the evolutions of data from the same family can be, up to final times
where the one parameter that differentiates them determines dispersal
or collapse.

Working with \texttt{bamps} and our AH locator enables the
classification as follows:
\begin{itemize}
\item At least down to a neighborhood of the threshold, \texttt{bamps}
  is able to evolve subcritical data until they disperse. We can hence
  confidently classify the data as subcritical just by looking at all
  fields dispersing at the end of each evolution. The blue line in
  Fig.~\ref{fig:illustrative} shows the field dispersal of a
  subcritical run.
\item Likewise in Fig.~\ref{fig:illustrative} one can see how the
  orange dashed line, corresponding to a supercritical run, blows up
  and stops before time~$20$. In the present evolutions we are not
  excising the trapped region, and so in the case of black hole
  formation, \texttt{bamps} can only evolve for a short time after
  trapped surface formation. We therefore look for AHs (see
  section~\ref{subsec:ahloc3d}) in the evolved data as a
  post-processing step. Only when we reliably find horizons do we
  classify the data as supercritical.
\end{itemize}

By categorizing the data in this way, we can estimate that the
threshold amplitude lies in the regime between the highest subcritical
amplitude and the lowest supercritical one. The process of classifying
the evolved initial data proceeds in stages, each increasing the
precision of our estimation of the threshold, thereby tuning closer
to~$A_{\star}$.

We employed a modified bisection method to reduce the size of the
interval that contains $A_\star$. We started the bisection with a
guess of trivially weak and strong data. After confirming the limits
of the threshold regime, we evolved data within it to further tune
those bounds at each stage. We increased the precision of these bounds
as far as our methods allowed. Assuming $A_\star\in[A_0,A_1]$, the
basic bisection method would proceed by choosing $A=(A_0+A_1)/2$,
performing an evolution to determine whether the data for $A$ is sub-
or supercritical, and adjusting the boundaries of the search interval
accordingly. In practice, we divided the threshold regime not
into~$2$, but typically into~$10$ subintervals. Such a
bracketed-search strategy can accelerate the search when performing
simulations for each $A$ in parallel. (Compare e.g.\ branch prediction
methods in CPU execution pipelines, where different if-clauses are
evaluated in parallel, but only the relevant result is used in the
end.) When successful, this method adds one decimal place per stage to
our limiting amplitudes. This approach furthermore has the advantage
that the data necessary for scaling-plots is already prepared directly
at the end of bisection without having to go back and resample.

%=========================================================
\subsection{Apparent Horizon Locator}
\label{subsec:ahloc3d}
%=========================================================

To conclusively classify a set of initial data as supercritical, the
AH finder \texttt{ahloc3d}~\cite{ahloc3d}, which is specifically
designed for use with \texttt{bamps} AMR data is used.
\texttt{ahloc3d} replaces the AH finder employed
in~\cite{HilWeyBru15,HilWeyBru17}, which had no functionality with AMR
data nor parallelization. During the development of~\texttt{ahloc3d}
thorough consistency checks were made with the results of the older
code.

\texttt{ahloc3d} uses a Strahlkörper representation to describe test
surfaces as
\begin{align}
  r &= h(\theta, \phi)
\end{align}
relative to a single center point, and evaluates the expansion
\begin{align}
  \Theta &= D_i s^i + \tensor{K}{_i_j} s^i s^j - K\,,
\end{align}
where~$s_i$ is the outward pointing normal vector on the test surface
and $K$ is the extrinsic curvature of the spacetime slice under
consideration, according to the algorithm described
in~\cite{Sch02}. Using two separate search methods, it first obtains a
rough estimate using a flow method using the expansion
flow~\cite{Tod91} to shrink a large test surface until an approximate
AH is found, i.e.~$\Theta$ satisfies a smallness condition.  It then
refines this estimate using a Newton-Raphson iteration to locate the
AH ($\Theta = 0$) with high accuracy. The search algorithm is fully
parallelized using MPI.  Its main limitation is the Strahlkörper
representation of the test surfaces, which makes it unable to find AHs
that are not radially convex. As we will discuss below, this has
prevented the further fine-tuning of several families of initial data,
as the AHs we found approached such non-convex shapes.

%=========================================================
\section{Numerical Results}
\label{sec:results}
%=========================================================

In this section we discuss the outcome of the bisection search
described in section~\ref{sec:phys_mod} for each of our six families.

%=========================================================
\subsection{Dynamics and threshold amplitudes}\label{subsec:threshold_amps}
%=========================================================

%=========================================================
\begin{figure}[t]
\includegraphics[scale=0.65]{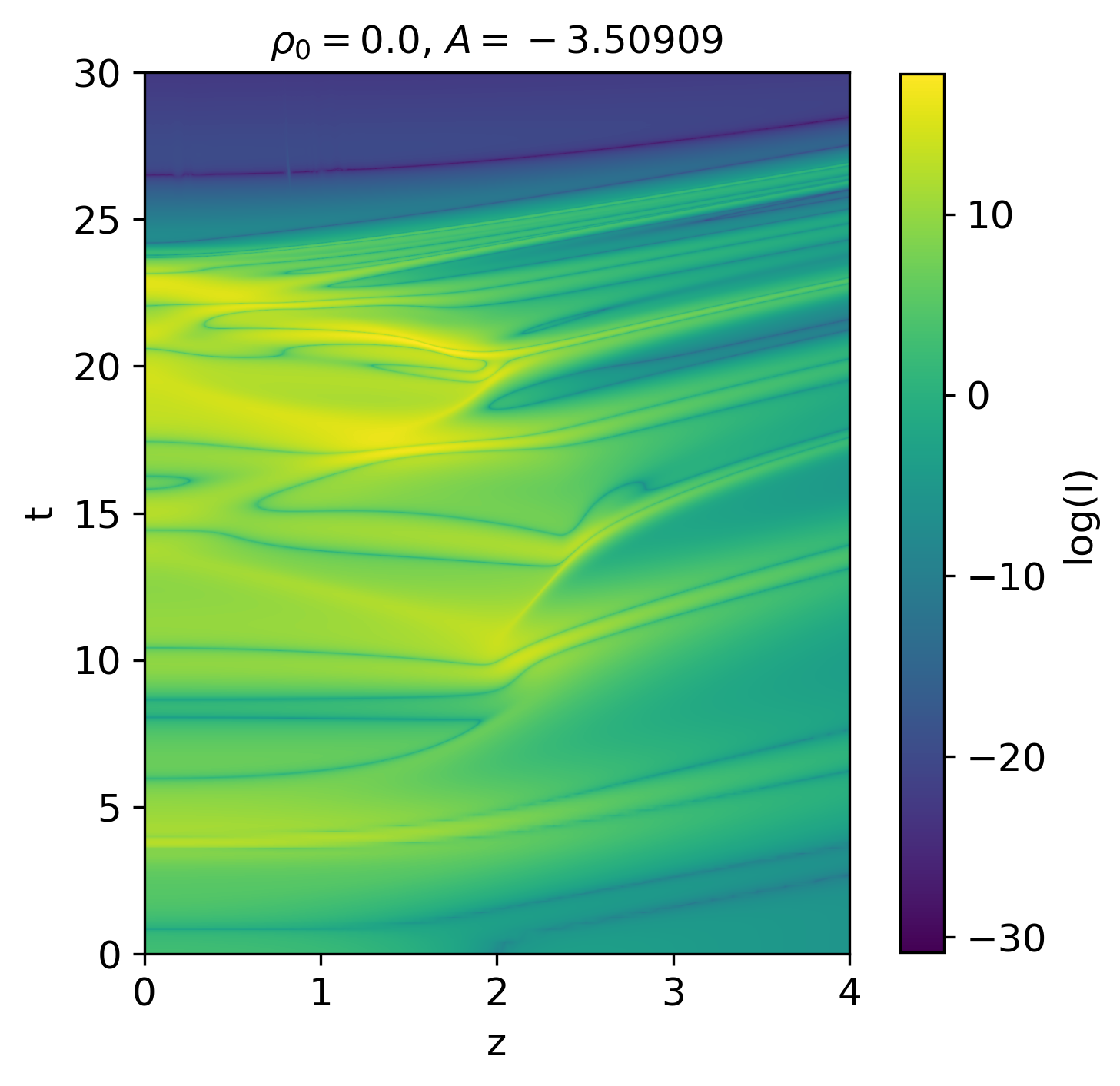}
\caption{In this figure a color map of the logarithm of the Kretschmann
  scalar along the symmetry axis and through time is shown for the
  oblate centered family $A < 0$, $\rho_0 = 0$, for the subcritical
  amplitude $A = -3.50909$. \label{fig:colormap}}
\end{figure}
%=========================================================

Basic evolution within each of the families is similar and follows the
behavior observed in~\cite{HilWeyBru17} for the centered~$A>0$ family.
Very weak data disperse rapidly, but as the strength of the data
increases the dynamics become more interesting. Considering the
Kretschmann scalar, the simplest non-vanishing curvature scalar in our
setup, we see wave-like propagation combined with larger spikes when
waves land on the symmetry axis and oscillate there. In comparison
with the centered data, the off-center families tend to form these
spikes further from the origin, probably because the curvature
propagates further in the~$\pm z$-directions before hitting the
symmetry axis. This difference would have made evolution of the
off-center data prohibitively expensive without AMR.  To illustrate
the dynamics qualitatively, in Fig.~\ref{fig:colormap} we show a 2D
plot of the logarithm of the Kretschmann scalar with respect to the
symmetry axis and coordinate time for a set of initial data close to
criticality ($A = -3.50909$) for the oblate ($A < 0$) centered
($\rho_0 = 0$) family. The region shown is relatively small in $z$,
but big enough to see that several peaks of the curvature scalar occur
away from the origin but along the symmetry axis before they
disperse. The results for all of the other families are similar, with
more and more off-centered peaks in the Kretschmann scalar appearing
as the threshold of black hole formation is approached.

There is a clear distinction between small initial data that lead to
dispersion and strong data that ends in the formation of a black
hole. The approach we use to classify the data (see
subsection~\ref{subsec:search}) is conservative but, given the
challenge these spacetimes pose and the unfortunate occasional
disagreement between different numerics, we feel it important to tread
lightly, compare carefully with the literature, and indeed to avoid
using another proxy for collapse. The results of our bisection search
are summarized in Table~\ref{tab:search_limits}, where we state the
highest subcritical and lowest supercritical amplitudes that we were
able to classify before \texttt{ahloc3d} failed to find apparent
horizons in near threshold simulations. These bounds are compatible
with the previous works~\cite{HilWeyBru17} and~\cite{LedKhi21} for the
centered families.

%=========================================================
\begin{table}[t!]
\begin{center}
\begin{tabular}{|c|c|c|c|}
\hline
 & $\rho_0$ & $A_{\text{sub}}$ & $A_{\text{sup}}$ \\
 \hline
 \multirow{3}{4em}{Prolate ($A > 0$)} & 0 & 4.69667  &  4.69680 \\
 & 4 & 0.09795 & 0.09870 \\
 & 5 & 0.0641 & 0.0645 \\
 \hline
  \multirow{3}{4em}{Oblate ($A < 0$)} & 0 & -3.50909  &  -3.50930 \\
 & 4 & -0.07546 & -0.07570 \\
 & 5 & -0.04878 & -0.04900 \\
 \hline
\end{tabular}
\end{center}
\caption{Results for the limits of the bisection search. The
  highest subcritical ($A_{\text{sub}}$) and lowest supercritical
  ($A_{\text{sup}}$) amplitudes for each family are displayed,
  defining the bounds of the threshold
  amplitudes.\label{tab:search_limits}}
\end{table}
%=========================================================

%=========================================================
\subsection{Apparent horizons}
%=========================================================

%=========================================================
\begin{figure}[t]
\includegraphics[scale=0.083]{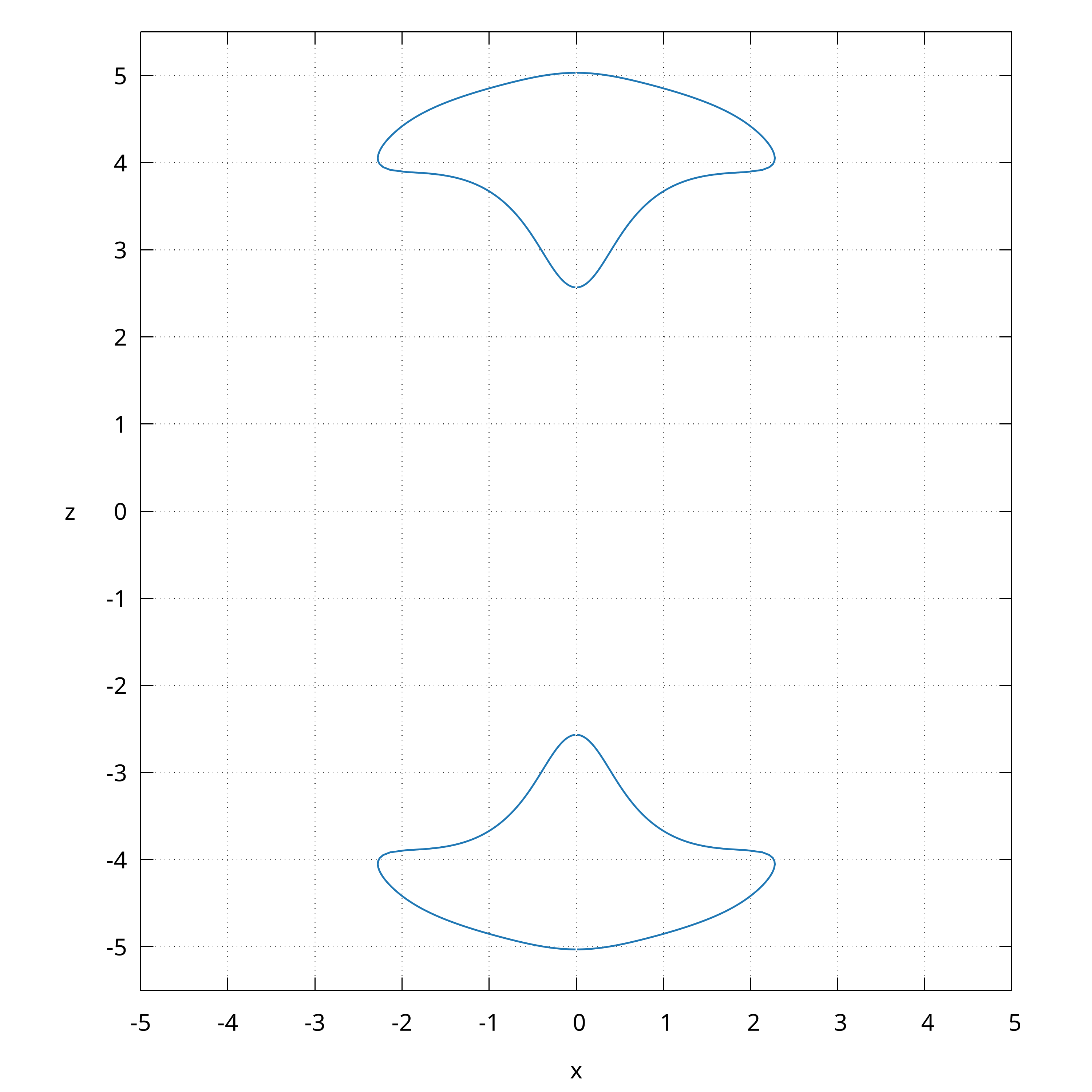}
\caption{A pair of apparent horizons found by \texttt{ahloc3d}. The
  horizontal axis represents the x direction and the vertical axis
  represents the symmetry axis, z. This is for the specific oblate
  off-centered family~$A<0$, $\rho_0=4$ with amplitude~$A = -0.07570$
  at time~$t \simeq 25.8$. Comparing, for example, with Fig.~$4$
  of~\cite{HilWeyBru17} we see that the apparent horizons are off-set
  by a greater coordinate distance.\label{fig:Horizon}}
\end{figure}
%=========================================================

Within a given (supercritical) member of a family, we reliably find
AHs after the first has appeared during evolution.  In agreement
with~\cite{HilWeyBru17, LedKhi21} for the centered prolate family,
close to the threshold we find that the AH for the off-centered oblate
families bifurcates (observe that this a statement in parameter space,
not within an individual spacetime). The AH forms approximately around
one of the large peaks in the Kretschmann scalar, along the symmetry
axis (forming a binary black hole spacetime if the event horizon shows
two components, but we do not investigate event horizons in this
work). In Fig.~\ref{fig:Horizon} we show this behavior for the oblate
($A < 0$), off-centered family~$\rho_0=4$ with amplitude~$A=-0.07570$.

Because of the reflection symmetry of the initial data, the two
horizons that appear are perfect copies, symmetric about the
equator. We do not have, however, explicit evidence from
\texttt{ahloc3d} that the remaining two families
(off-center~$\rho_0=4,5$, prolate~$A>0$) also have bifurcated,
two-component AHs. To find these, this post-processing software needs
to be run close enough to the time of formation of the AH, as shortly
after trapped surface formation the underlying numerical evolution
will fail. This makes it difficult to hit the right time to
search. Despite the lack of explicit evidence, we have strong
indications that this is a common feature for all six of the families
we studied. Close to the threshold the global peak of the maximum of
the Kretschmann scalar always occurs, both for subcritical and
supercritical cases, away from the origin and along the symmetry
axis. Furthermore, when facing the difficulties with \texttt{ahloc3d}
we found some abnormal shapes that are not trustworthy as AH but that
seem to indicate that the formation of disjoint AHs is about to occur,
or that the has already happened, and that the algorithm of the
software needs improvement. Presumably, if we chose initial data
without reflection symmetry, and tuned well enough to the threshold,
an AH would form only on one side of the equator. Both of the latter
two points need further work.

%=========================================================
\subsection{Scaling}
%=========================================================

%=========================================================
\begin{figure}[t]
\includegraphics[scale=0.56]{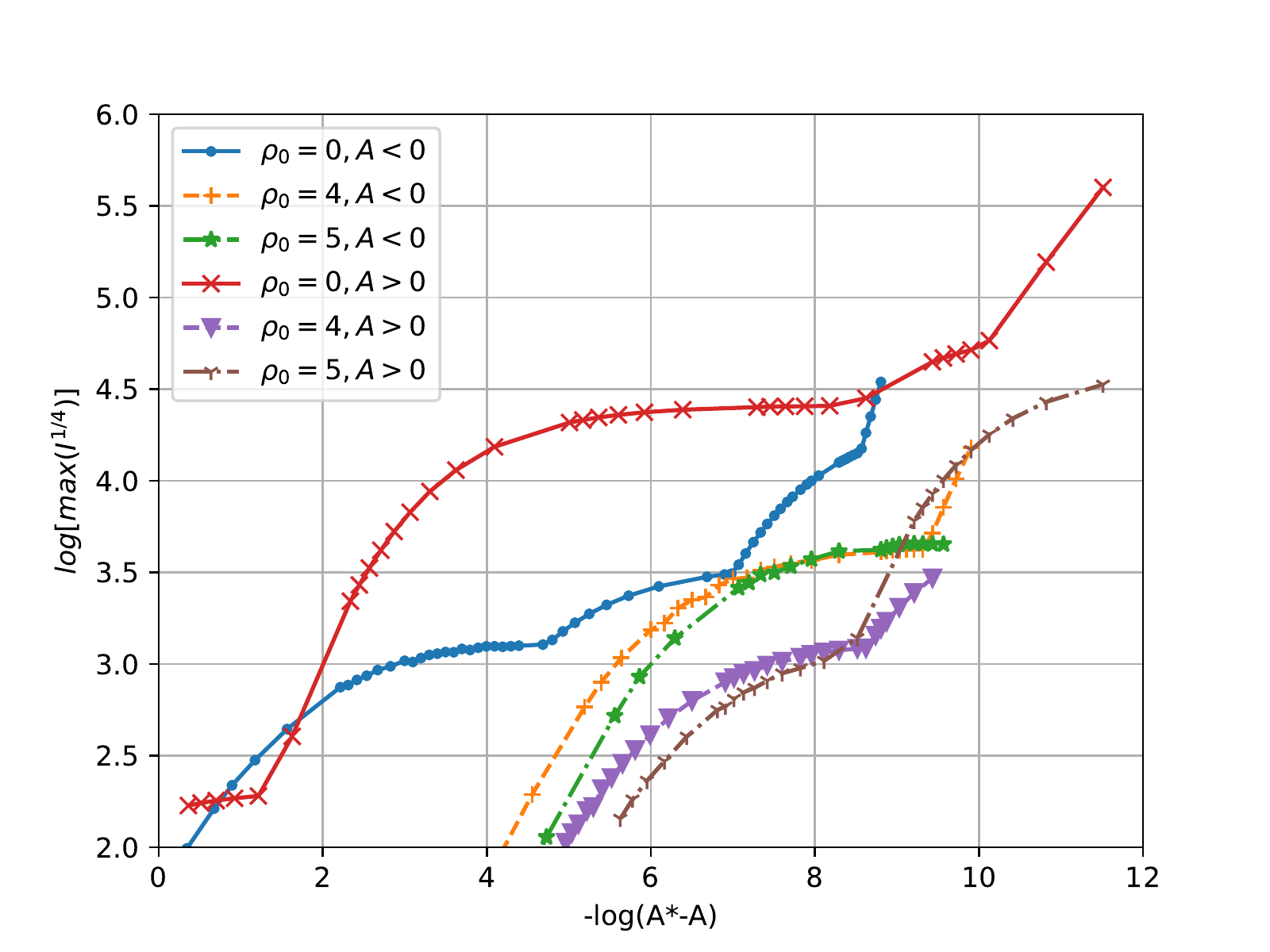}
\caption{The scaling of the Kretschmann scalar for all the studied
  families is shown. In the horizontal axis we represent the
  logarithmic distance to the critical point and in the vertical axis
  the logarithm of the maximum of the Kretschmann as~$I^{(1/4)}$.
  The red curve compares with Fig.~$4$
  of~\cite{HilWeyBru17} and the blue with the green in Fig.~$1$
  of~\cite{LedKhi21}.
 \label{fig:scaling}}
\end{figure}
%=========================================================

According to~\cite{GarDun98}, if critical phenomena are present in the
collapse of gravitational waves, any curvature scalar invariant should
show universal power-law scaling in the subcritical regime. As we are
working in vacuum the Ricci scalar vanishes, so we investigate the
scaling behavior of the
Kretschmann scalar. In Fig.~\ref{fig:scaling} we plot
the logarithm of the maximum of
the Kretschmann scalar in the form~$I^{1/4}$ against the logarithmic distance
to the critical amplitude.
(The exponent $\frac{1}{4}$ is chosen to obtain units of one over length.)
First, it is clear that the maximum value
that the Kretschmann scalar attains depends on the amplitude of the
initial data for each family. Second, for each individual family this
result is compatible with the power-law behavior in
equation~\eqref{eq:I_power_law}, and furthermore agrees very well
with~\cite{HilWeyBru17, LedKhi21} for the two centered families.
There is, however, no evidence of a universal exponent. A priori, this
result is compatible with each family having a different exponent,
but, extending these lines further to the right (which is very
computationally expensive and challenging) would be necessary to make
a conclusive statement about universality. If the exponents were truly
different, we could wonder whether there exists a finite number of
such exponents, leading to a new paradigm of universality, or else, if
there are simply as many exponents as families of initial
data. Evidently more investigation is needed. Referring to DSS, only
one family (the blue curve in Fig.~\ref{fig:scaling}) shows enough
periods for us to claim that it behaves as \textit{approximately}-DSS
and more periods are needed in the other families. This is also the
reason why we postpone treating the errors and interpret this result
as qualitative for now.

%=========================================================
\subsection{Echoes and universality}
%=========================================================

%=========================================================
\begin{figure}[t]
\includegraphics[scale=0.5]{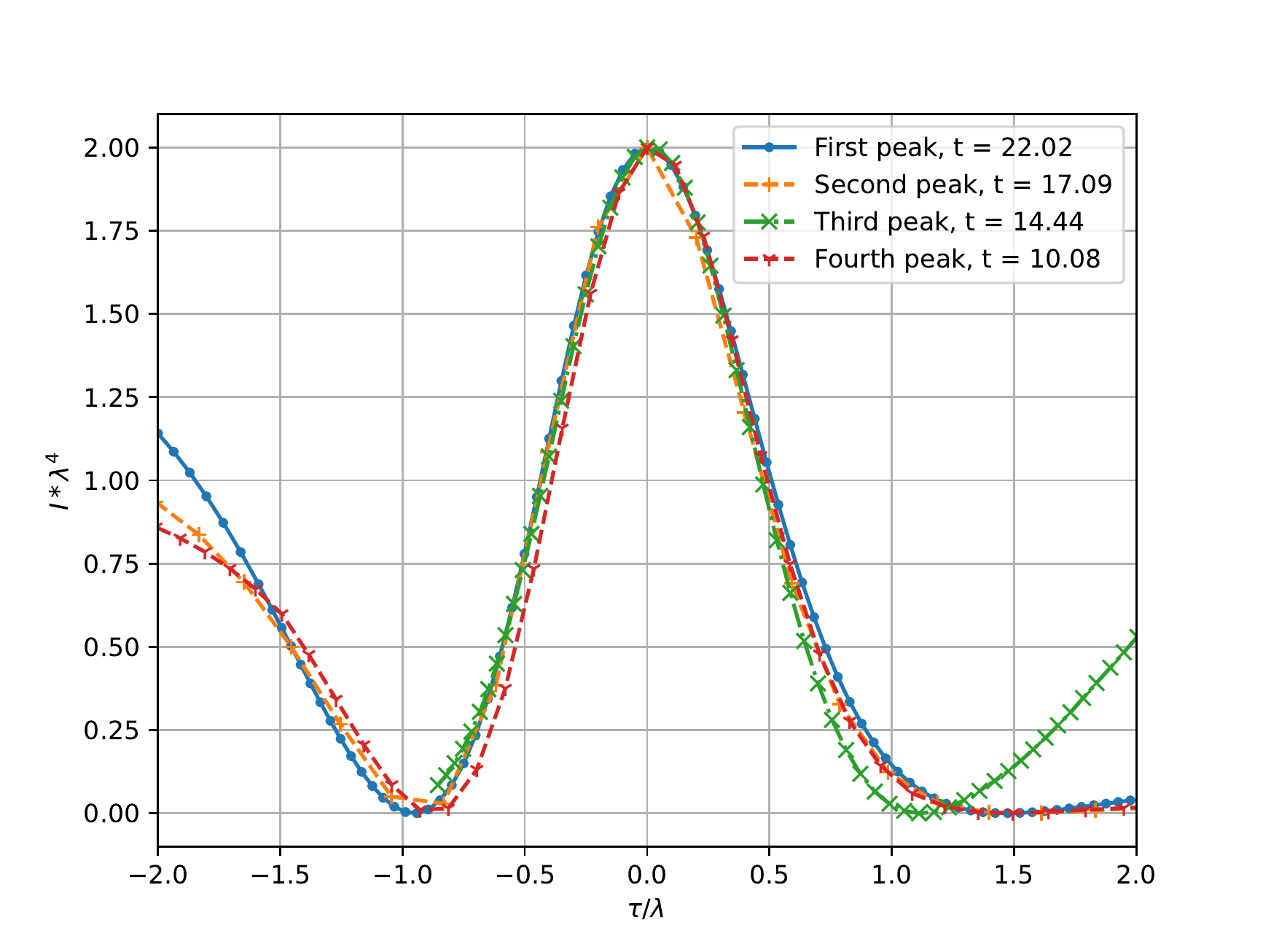}
\caption{Shown are four echoes for the family of initial
  data $A<0$, $\rho_0 = 0$ with $A = -3.50909$. Each echo corresponds
  to a peak of the maximum of the Kretschmann scalar at
  different times against the proper time. Both axes are rescaled
  for each curve by the same constant~$\lambda$ which is chosen such that the
  maximum will correspond to~$2$ in the plot. The largest ratio
  of~$\lambda$'s across such curves
  is~$\sim3$. \label{fig:echoes_rho0_neg}}
\end{figure}
%=========================================================

%=========================================================
\begin{figure}[t!]
\includegraphics[scale=0.55]{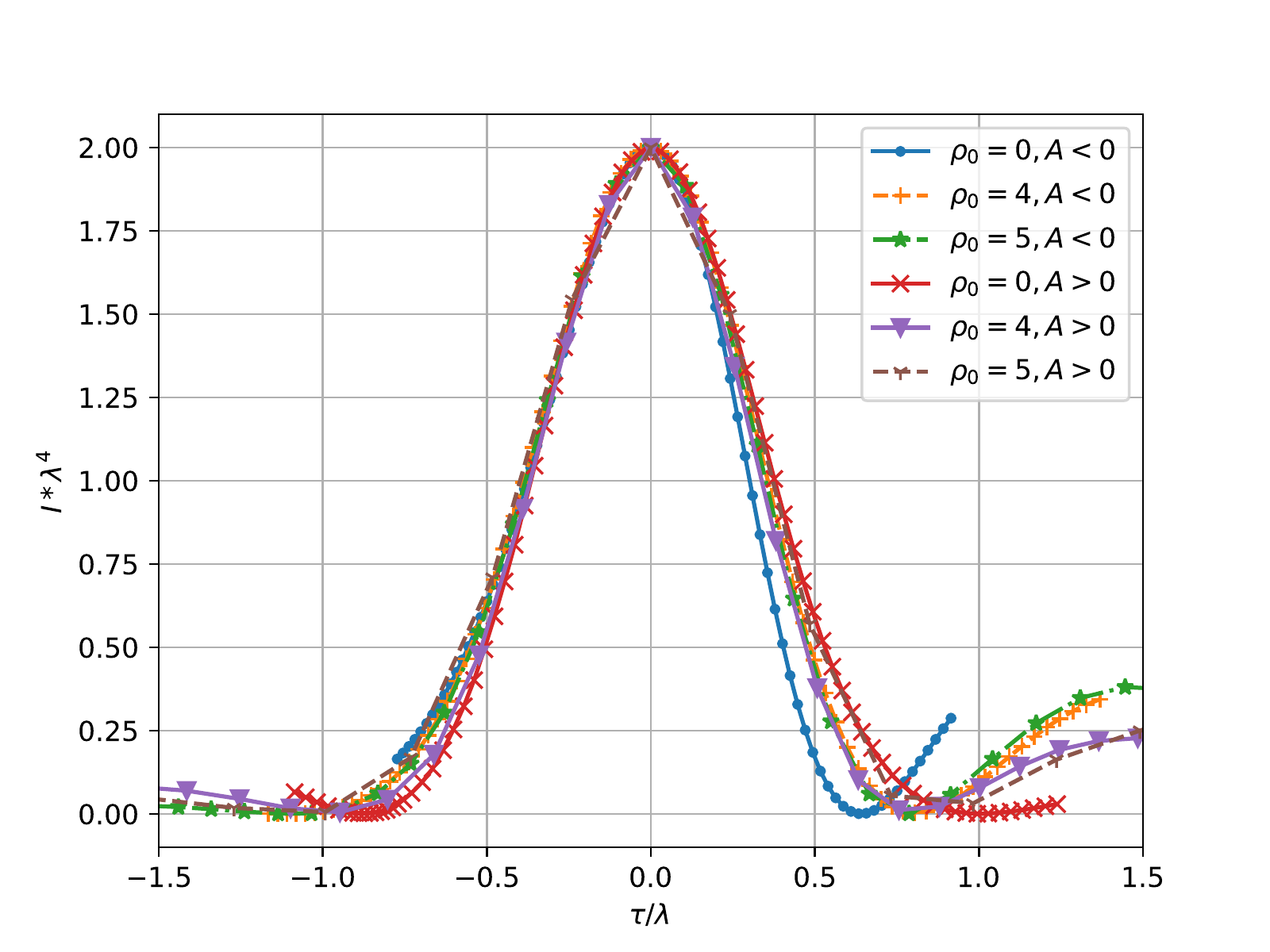}
\caption{In this figure we show the largest peak of the Kretschmann
  scalar, for the highest subcritical amplitude of Table
  \ref{tab:search_limits}, for each of the six studied families,
  against the proper time along the integral curve of~$n^a$. As in
  Fig.~\ref{fig:echoes_rho0_neg}, both the proper time and Kretschmann
  scalar are rescaled by a constant~$\lambda$ chosen such that the
  maximum of the Kretschmann scalar occurs at 2 in the plot. For these
  curves the largest ratio of~$\lambda$'s is around~$10$, corresponding
  to a ratio~$10^4$ in the peak of the Kretschmann itself. The data
  for the prolate ($A > 0$) off-centered families and for the oblate
  ($A < 0$) centered family have been flipped along the horizontal axis
  for a better match. \label{fig:echoes_all}}
\end{figure}
%=========================================================

In the case of a massless scalar field collapse in spherical symmetry
\cite{Cho93}, the critical solution shows DSS behavior, however, in
the axisymmetric collapse of gravitational waves in vacuum the picture
is more complicated. As discussed above, we observe that for a near
threshold evolution~$A \lesssim A_\star$ (flipped for~$A<0$ families),
several large local peaks in the Kretschmann scalar appear along the
symmetry axis before the data eventually disperse. In
Fig.~\ref{fig:echoes_rho0_neg} we plot these peaks against proper time
along timelike curves (the integral curves of the unit normal
vector~$n^a$) that pass through the maxima (one curve per peak), and
compare their shape by rescaling them with a constant~$\lambda$ so
that we plot dimensionless quantities on both axes. In the figure, we
take the oblate ($A < 0$) centered family~$\rho_0=0$. We allow
ourselves the freedom to flip the axis for individual curves to take
care of features propagating up or down the~$z$-axis. The agreement of
the four curves is striking, especially given that the values
of~$\lambda$ for each of the curves vary by a factor of~$\sim3$,
corresponding to a little less than two orders of magnitude in the
Kretschmann scalar itself. Four echoes were found for four different
times, in qualitative agreement with~\cite{LedKhi21} where they show
in Fig.~3 a similar plot for a Teukolsky wave. It is important to
remark that this feature is exclusive to near threshold evolutions. In
contrast, for small initial data only one peak of the Kretschmann
scalar occurs, and no such repeated features appear. Analogous results
were found in all of the other families, however, the echoing is not
as clean as for the displayed family. Given that~$A-A_\star$ has no
absolute meaning, one might argue that for the other families our data
are not close enough to the threshold solution to show such good
results. On the symmetry axis, the scalar quantity studied by Ledvinka
and Khirnov in~\cite{LedKhi21} is directly related to the Kretschmann
scalar. One might expect their profiles to agree with ours, but, the
normal vector associated with the foliation of~\cite{LedKhi21} does
not coincide with ours, and thus the integral curves along which we
plot do not either. We thus postpone a detailed comparison. We have
not identified clearly whether these repeated features correspond to
true DSS behavior. In our coordinates they appear without regular time
intervals, and our curvature scaling plots indicate neither universal
power-laws nor uniformly periodic wiggles, so we have no reason to
expect true DSS. That said, this is clear evidence of phenomenology
familiar from the spherical setting carrying over.

In order to compare the spacetime behavior among families, in
Fig.~\ref{fig:echoes_all} we plot the profile of the Kretschmann
scalar against the proper time along a timelike curve that passes
through the maxima, for near critical data as done in
Fig.~\ref{fig:echoes_rho0_neg}, but for the largest peak within our
best-tuned data within each different family. Each of these lines
corresponds to the most right placed point of Fig.~\ref{fig:scaling}
for each family, being as close to the threshold solution as
possible. Again the shapes around the peaks agree, showing a common
feature that appears when evolving initial data with amplitude~$A
\simeq A_\star$ independently of the chosen family of initial data,
and again, in concordance with~\cite{LedKhi21}.

%=========================================================
\section{Summary and discussion}
\label{sec:summary}
%=========================================================

Earlier work~\cite{HilWeyBru17} on gravitational wave collapse
with~\texttt{bamps} was severely hampered by the rapidly increasing
computational cost near the threshold. With an allocation of
around~$10$ million core hours, in that study we were able to tune
just a single family to the threshold of collapse. For that
reason~\texttt{bamps} has undergone a major overhaul, and now fully
supports adaptive mesh-refinement, the details of which will appear in
a stand-alone report. Presently, with a similar allocation we were
able to tune six families to within a comparable distance to the
threshold of collapse. What is more, it is expected that as we now
push even closer to the threshold and progressively finer spacetime
features appear, this improvement in efficiency will become ever more
pronounced.

The obvious question is then, with this additional efficiency why have
we not already pushed one (or more) of our families closer to the
threshold? The answer is that we still encounter difficulties in the
bisection, and in particular in classifying spacetimes close to the
threshold. There are two main reasons for this; first, we cannot
preclude the formation of coordinate singularities, which no amount of
AMR could cure, and which would prevent classification if they appear
before trapped surface formation. The choice of the gauge
parameter~$\eta_S$ as defined in~\eqref{eqn:Gauge_Sources} for
instance appears to be highly important. Related to this is the
appearance of constraint violation, which does get worse in the
strong-field regime and especially in the presence of fine features
(but which we believe is not the leading problem in our present
data). Unsurprisingly we found that our evolutions were more
numerically challenging close to the threshold. Constraint violation
does get worse near the threshold. Usually the constraint violation
stays in a range of about~$10^{-8}$ to~$10^{-6}$ for small initial
data. However, when the Kretschmann scalar grows this violation might
reach up to~$10^{-3}$, but still remains several orders of magnitude
smaller than the relevant evolved quantities. The second issue we
encountered was that, particularly for the off-centered families, the
shapes of the apparent horizons change drastically the closer we get
to the threshold. We suspect that there may be AHs which cannot be
captured by \texttt{ahloc3d} due to the Strahlkörper representation
described in section~\ref{subsec:ahloc3d}. Fortunately, strategies are
available to avoid both of these issues in the future.

Despite these shortcomings, our setup allows careful examination of
the threshold of black hole formation of axisymmetric gravitational
waves, which is, by default, beyond spherical symmetry in~$3+1$
spacetime dimensions. Close to the threshold, we find explicitly that
three of our six families form two disjoint apparent horizons. In the
remaining three families we expect the same because the spikes in the
Kretschmann scalar form away from the origin. These spikes are further
separated for off-center data than for centered data. In hindsight,
this was perhaps foreseeable because, since our seed functions were
off-set in the cylindrical polar direction~$\rho$, the waves have time
to propagate in the~$z$-direction before they hit the symmetry
axis. It would be interesting to evolve initial data families off-set
in both~$\rho$ and~$z$ to examine how generic the appearance of pairs
of AHs is within families of reflection symmetric data. With the
(constraint solved) Teukolsky waves of~\cite{LedKhi21} and the present
work, this does however appear a fairly robust feature.

Our results for the scaling of the Kretschmann scalar shown in
Fig.~\ref{fig:scaling} agree perfectly
with~\cite{HilWeyBru17,LedKhi21}, making us confident that our code is
reliable as we were able to reproduce results for the centered Brill
waves. Our main objective was to help establish the extent to which
the standard picture of critical collapse extends beyond spherical
symmetry. As discussed in the introduction, a number of studies
suggest that this story is a subtle one, and the present work further
reinforces this perspective. We examined four families of off-center
initial data for the first time. Considering the scaling plot for each
family individually, it is tempting to argue that the data take the
form of a power-law plus a possibly periodic wiggle. But at least to
the level of tuning we have presented here, there appears to be
neither a universal power nor period in the wiggle. It is possible
that we simply need greater numerical accuracy, but given the solid
agreement with the independent implementation in~\cite{LedKhi21} that
does seem unlikely. The evidence at hand thus suggests that the
exponents of the respective power-laws and periods of the wiggles, if
the latter can even be defined, are family dependent. Presumably, this
corresponds to the manifestation of different threshold solutions, and
so departs from the standard picture in spherical symmetry. Assuming
this is the case, a natural question is whether threshold solutions
can be well described by a finite, countable, or uncountable number of
parameters. Much more work is needed to shed light on this question.

Most interestingly, our evidence strongly suggests that aspects of
behavior familiar from the spherical setting do carry over. Looking at
Fig.~\ref{fig:echoes_all}, it is clear that, close enough to the
threshold solution, all six of our families present strikingly similar
behavior for the Kretschmann scalar, with a peak of practically the
same shape appearing at different scales. This is a clear indication
that some kind of universality still remains. This is furthermore in
good agreement with the findings by~\cite{LedKhi21} for alternative
families of initial data. Likewise we also find, within individual
families, that repeated echoes appear in the curvature scalar.

There are a number of obvious ways in which to extend our work. First,
improvements to \texttt{ahloc3d} are needed so that the apparent
horizons that do not follow the Strahlkörper parametrization can be
found (see~\cite{Khi21}, where the same problem was faced, for a
possible solution). It could be possible to analyze even the sets of
data that were currently produced within this work to push these same
bisection searches further. Next, is the treatment of alternative
families, including radially off-set Brill waves and (constraint
solved) Teukolsky waves. We also expect that finer control of
constraint violation and coordinates would be of benefit. For the
latter we have already implemented the DF-GHG
formulation~\cite{Hil15,HilHarBug16,BhaHilRaj21}, but the complete
generalization of the outer boundary conditions to that setting is
ongoing. A further question, already mentioned in passing
in~\cite{HilWeyBru17}, is whether or not the curvature spikes we
observe have anything to do with BKL behavior. As such,
following~\cite{GarPre20}, it would be interesting to calculate the
expected behavior of the Kretschmann scalar for
comparison. Ultimately, as \texttt{bamps} undergoes further
development, we aim to relax our symmetry assumptions, to make full 3d
evolutions at the threshold of vacuum collapse possible. Progress on
all of these fronts will be reported elsewhere.

%=========================================================
\acknowledgments
%=========================================================

We are grateful for the computational resources provided by the
Leibniz Supercomputing Centre (LRZ) [project~pn34vo], without which
this work would have been impossible. The work was partially supported
by the FCT (Portugal) IF Program~IF/00577/2015,
PTDC/MAT-APL/30043/2017, IDPASC programs PD/BD/135434/2017 and
COVID/BD/152485/2022, and Project~No.~UIDB/00099/2020; also partially
supported by the Deutsche Forschungsgemeinschaft (DFG) under Grant
No. 406116891 within the Research Training Group RTG 2522/1.

%=========================================================
\bibliography{EBW.bbl}

\begin{thebibliography}{50}%
\makeatletter
\providecommand \@ifxundefined [1]{%
 \@ifx{#1\undefined}
}%
\providecommand \@ifnum [1]{%
 \ifnum #1\expandafter \@firstoftwo
 \else \expandafter \@secondoftwo
 \fi
}%
\providecommand \@ifx [1]{%
 \ifx #1\expandafter \@firstoftwo
 \else \expandafter \@secondoftwo
 \fi
}%
\providecommand \natexlab [1]{#1}%
\providecommand \enquote  [1]{``#1''}%
\providecommand \bibnamefont  [1]{#1}%
\providecommand \bibfnamefont [1]{#1}%
\providecommand \citenamefont [1]{#1}%
\providecommand \href@noop [0]{\@secondoftwo}%
\providecommand \href [0]{\begingroup \@sanitize@url \@href}%
\providecommand \@href[1]{\@@startlink{#1}\@@href}%
\providecommand \@@href[1]{\endgroup#1\@@endlink}%
\providecommand \@sanitize@url [0]{\catcode `\\12\catcode `\$12\catcode
  `\&12\catcode `\#12\catcode `\^12\catcode `\_12\catcode `\%12\relax}%
\providecommand \@@startlink[1]{}%
\providecommand \@@endlink[0]{}%
\providecommand \url  [0]{\begingroup\@sanitize@url \@url }%
\providecommand \@url [1]{\endgroup\@href {#1}{\urlprefix }}%
\providecommand \urlprefix  [0]{URL }%
\providecommand \Eprint [0]{\href }%
\providecommand \doibase [0]{https://doi.org/}%
\providecommand \selectlanguage [0]{\@gobble}%
\providecommand \bibinfo  [0]{\@secondoftwo}%
\providecommand \bibfield  [0]{\@secondoftwo}%
\providecommand \translation [1]{[#1]}%
\providecommand \BibitemOpen [0]{}%
\providecommand \bibitemStop [0]{}%
\providecommand \bibitemNoStop [0]{.\EOS\space}%
\providecommand \EOS [0]{\spacefactor3000\relax}%
\providecommand \BibitemShut  [1]{\csname bibitem#1\endcsname}%
\let\auto@bib@innerbib\@empty
%</preamble>
\bibitem [{\citenamefont {Choptuik}(1993)}]{Cho93}%
  \BibitemOpen
  \bibfield  {author} {\bibinfo {author} {\bibfnamefont {M.~W.}\ \bibnamefont
  {Choptuik}},\ }\bibfield  {title} {\bibinfo {title} {Universality and scaling
  in gravitational collapse of massless scalar field},\ }\href@noop {}
  {\bibfield  {journal} {\bibinfo  {journal} {Phys. Rev. Lett.}\ }\textbf
  {\bibinfo {volume} {70}},\ \bibinfo {pages} {9} (\bibinfo {year}
  {1993})}\BibitemShut {NoStop}%
\bibitem [{\citenamefont {Reiterer}\ and\ \citenamefont
  {Trubowitz}(2019)}]{ReiTru19}%
  \BibitemOpen
  \bibfield  {author} {\bibinfo {author} {\bibfnamefont {M.}~\bibnamefont
  {Reiterer}}\ and\ \bibinfo {author} {\bibfnamefont {E.}~\bibnamefont
  {Trubowitz}},\ }\bibfield  {title} {\bibinfo {title} {{Choptuik’s Critical
  Spacetime Exists}},\ }\href {https://doi.org/10.1007/s00220-019-03413-8}
  {\bibfield  {journal} {\bibinfo  {journal} {Commun. Math. Phys.}\ }\textbf
  {\bibinfo {volume} {368}},\ \bibinfo {pages} {143} (\bibinfo {year}
  {2019})}\BibitemShut {NoStop}%
%%CITATION = CMPHA,368,143;%%
\bibitem [{\citenamefont {Garfinkle}\ and\ \citenamefont
  {Duncan}(1998)}]{GarDun98}%
  \BibitemOpen
  \bibfield  {author} {\bibinfo {author} {\bibfnamefont {D.}~\bibnamefont
  {Garfinkle}}\ and\ \bibinfo {author} {\bibfnamefont {G.~C.}\ \bibnamefont
  {Duncan}},\ }\bibfield  {title} {\bibinfo {title} {{Scaling of curvature in
  subcritical gravitational collapse}},\ }\href
  {https://doi.org/10.1103/PhysRevD.58.064024} {\bibfield  {journal} {\bibinfo
  {journal} {Phys.Rev.}\ }\textbf {\bibinfo {volume} {D58}},\ \bibinfo {pages}
  {064024} (\bibinfo {year} {1998})},\ \Eprint
  {https://arxiv.org/abs/gr-qc/9802061} {arXiv:gr-qc/9802061 [gr-qc]}
  \BibitemShut {NoStop}%
%%CITATION = GR-QC/9802061;%%
\bibitem [{\citenamefont {Gundlach}(1998)}]{Gun97}%
  \BibitemOpen
  \bibfield  {author} {\bibinfo {author} {\bibfnamefont {C.}~\bibnamefont
  {Gundlach}},\ }\bibfield  {title} {\bibinfo {title} {Pseudo-spectral apparent
  horizon finders: An efficient new algorithm},\ }\href@noop {} {\bibfield
  {journal} {\bibinfo  {journal} {Phys. Rev. D}\ }\textbf {\bibinfo {volume}
  {57}},\ \bibinfo {pages} {863} (\bibinfo {year} {1998})},\ \Eprint
  {https://arxiv.org/abs/gr-qc/9707050} {gr-qc/9707050} \BibitemShut {NoStop}%
\bibitem [{\citenamefont {Hod}\ and\ \citenamefont {Piran}(1997)}]{HodPir97}%
  \BibitemOpen
  \bibfield  {author} {\bibinfo {author} {\bibfnamefont {S.}~\bibnamefont
  {Hod}}\ and\ \bibinfo {author} {\bibfnamefont {T.}~\bibnamefont {Piran}},\
  }\bibfield  {title} {\bibinfo {title} {{Fine structure of Choptuik's mass
  scaling relation}},\ }\href {https://doi.org/10.1103/PhysRevD.55.R440}
  {\bibfield  {journal} {\bibinfo  {journal} {Phys.\ Rev.\ D}\ }\textbf
  {\bibinfo {volume} {55}},\ \bibinfo {pages} {440} (\bibinfo {year} {1997})},\
  \Eprint {https://arxiv.org/abs/gr-qc/9606087} {arXiv:gr-qc/9606087}
  \BibitemShut {NoStop}%
\bibitem [{\citenamefont {Gundlach}\ and\ \citenamefont
  {Mart{\'i}n-Garc{\'i}a}(2007)}]{GunGar07}%
  \BibitemOpen
  \bibfield  {author} {\bibinfo {author} {\bibfnamefont {C.}~\bibnamefont
  {Gundlach}}\ and\ \bibinfo {author} {\bibfnamefont {J.~M.}\ \bibnamefont
  {Mart{\'i}n-Garc{\'i}a}},\ }\bibfield  {title} {\bibinfo {title} {Critical
  phenoena in gravitational collapse},\ }\bibfield  {journal} {\bibinfo
  {journal} {Living Reviews in Relativity}\ }\textbf {\bibinfo {volume} {10}},\
  \href {https://doi.org/10.12942/lrr-2007-5} {10.12942/lrr-2007-5} (\bibinfo
  {year} {2007})\BibitemShut {NoStop}%
\bibitem [{\citenamefont {Martin-Garcia}\ and\ \citenamefont
  {Gundlach}(1999)}]{GarGun98}%
  \BibitemOpen
  \bibfield  {author} {\bibinfo {author} {\bibfnamefont {J.~M.}\ \bibnamefont
  {Martin-Garcia}}\ and\ \bibinfo {author} {\bibfnamefont {C.}~\bibnamefont
  {Gundlach}},\ }\bibfield  {title} {\bibinfo {title} {All nonspherical
  perturbations of the choptuik spacetime decay},\ }\href@noop {} {\bibfield
  {journal} {\bibinfo  {journal} {Phys. Rev. D}\ }\textbf {\bibinfo {volume}
  {59}},\ \bibinfo {pages} {064031} (\bibinfo {year} {1999})},\ \Eprint
  {https://arxiv.org/abs/gr-qc/9809059} {gr-qc/9809059} \BibitemShut {NoStop}%
\bibitem [{\citenamefont {Choptuik}\ \emph {et~al.}(2003)\citenamefont
  {Choptuik}, \citenamefont {Hirschmann}, \citenamefont {Liebling},\ and\
  \citenamefont {Pretorius}}]{ChoHirLie03}%
  \BibitemOpen
  \bibfield  {author} {\bibinfo {author} {\bibfnamefont {M.~W.}\ \bibnamefont
  {Choptuik}}, \bibinfo {author} {\bibfnamefont {E.~W.}\ \bibnamefont
  {Hirschmann}}, \bibinfo {author} {\bibfnamefont {S.~L.}\ \bibnamefont
  {Liebling}},\ and\ \bibinfo {author} {\bibfnamefont {F.}~\bibnamefont
  {Pretorius}},\ }\bibfield  {title} {\bibinfo {title} {Critical collapse of
  the massless scalar field in axisymmetry},\ }\href@noop {} {\bibfield
  {journal} {\bibinfo  {journal} {Phys. Rev. D}\ }\textbf {\bibinfo {volume}
  {68}},\ \bibinfo {pages} {044007} (\bibinfo {year} {2003})},\ \Eprint
  {https://arxiv.org/abs/gr-qc/0305003} {gr-qc/0305003} \BibitemShut {NoStop}%
\bibitem [{\citenamefont {Baumgarte}(2018)}]{Bau18}%
  \BibitemOpen
  \bibfield  {author} {\bibinfo {author} {\bibfnamefont {T.~W.}\ \bibnamefont
  {Baumgarte}},\ }\bibfield  {title} {\bibinfo {title} {{Aspherical
  deformations of the Choptuik spacetime}},\ }\href@noop {} {\  (\bibinfo
  {year} {2018})},\ \Eprint {https://arxiv.org/abs/1807.10342}
  {arXiv:1807.10342 [gr-qc]} \BibitemShut {NoStop}%
%%CITATION = ARXIV:1807.10342;%%
\bibitem [{\citenamefont {Healy}\ and\ \citenamefont
  {Laguna}(2014)}]{HeaLag13}%
  \BibitemOpen
  \bibfield  {author} {\bibinfo {author} {\bibfnamefont {J.}~\bibnamefont
  {Healy}}\ and\ \bibinfo {author} {\bibfnamefont {P.}~\bibnamefont {Laguna}},\
  }\bibfield  {title} {\bibinfo {title} {{Critical Collapse of Scalar Fields
  Beyond Axisymmetry}},\ }\href@noop {} {\bibfield  {journal} {\bibinfo
  {journal} {Gen. Rel. Grav.}\ }\textbf {\bibinfo {volume} {46}},\ \bibinfo
  {pages} {1722} (\bibinfo {year} {2014})},\ \Eprint
  {https://arxiv.org/abs/1310.1955} {arXiv:1310.1955 [gr-qc]} \BibitemShut
  {NoStop}%
%%CITATION = ARXIV:1310.1955;%%
\bibitem [{\citenamefont {Deppe}\ \emph {et~al.}(2018)\citenamefont {Deppe},
  \citenamefont {Kidder}, \citenamefont {Scheel},\ and\ \citenamefont
  {Teukolsky}}]{DepKidSch18}%
  \BibitemOpen
  \bibfield  {author} {\bibinfo {author} {\bibfnamefont {N.}~\bibnamefont
  {Deppe}}, \bibinfo {author} {\bibfnamefont {L.~E.}\ \bibnamefont {Kidder}},
  \bibinfo {author} {\bibfnamefont {M.~A.}\ \bibnamefont {Scheel}},\ and\
  \bibinfo {author} {\bibfnamefont {S.~A.}\ \bibnamefont {Teukolsky}},\
  }\bibfield  {title} {\bibinfo {title} {{Critical behavior in 3-d
  gravitational collapse of massless scalar fields}},\ }\href@noop {} {\
  (\bibinfo {year} {2018})},\ \Eprint {https://arxiv.org/abs/1802.08682}
  {arXiv:1802.08682 [gr-qc]} \BibitemShut {NoStop}%
%%CITATION = ARXIV:1802.08682;%%
\bibitem [{\citenamefont {Baumgarte}\ \emph {et~al.}(2019)\citenamefont
  {Baumgarte}, \citenamefont {Gundlach},\ and\ \citenamefont
  {Hilditch}}]{BauGunHil19}%
  \BibitemOpen
  \bibfield  {author} {\bibinfo {author} {\bibfnamefont {T.~W.}\ \bibnamefont
  {Baumgarte}}, \bibinfo {author} {\bibfnamefont {C.}~\bibnamefont
  {Gundlach}},\ and\ \bibinfo {author} {\bibfnamefont {D.}~\bibnamefont
  {Hilditch}},\ }\bibfield  {title} {\bibinfo {title} {{Critical phenomena in
  the gravitational collapse of electromagnetic waves}},\ }\href
  {https://doi.org/10.1103/PhysRevLett.123.171103} {\bibfield  {journal}
  {\bibinfo  {journal} {Phys. Rev. Lett.}\ }\textbf {\bibinfo {volume} {123}},\
  \bibinfo {pages} {171103} (\bibinfo {year} {2019})},\ \Eprint
  {https://arxiv.org/abs/1909.00850} {arXiv:1909.00850 [gr-qc]} \BibitemShut
  {NoStop}%
\bibitem [{\citenamefont {Perez~Mendoza}\ and\ \citenamefont
  {Baumgarte}(2021)}]{MenBau21}%
  \BibitemOpen
  \bibfield  {author} {\bibinfo {author} {\bibfnamefont {M.~F.}\ \bibnamefont
  {Perez~Mendoza}}\ and\ \bibinfo {author} {\bibfnamefont {T.~W.}\ \bibnamefont
  {Baumgarte}},\ }\bibfield  {title} {\bibinfo {title} {Critical phenomena in
  the gravitational collapse of electromagnetic dipole and quadrupole waves},\
  }\href {https://doi.org/10.1103/PhysRevD.103.124048} {\bibfield  {journal}
  {\bibinfo  {journal} {Phys. Rev. D}\ }\textbf {\bibinfo {volume} {103}},\
  \bibinfo {pages} {124048} (\bibinfo {year} {2021})}\BibitemShut {NoStop}%
\bibitem [{\citenamefont {Su\'arez~Fern\'andez}\ \emph
  {et~al.}(2021{\natexlab{a}})\citenamefont {Su\'arez~Fern\'andez},
  \citenamefont {Vicente},\ and\ \citenamefont {Hilditch}}]{SuaVicHil21}%
  \BibitemOpen
  \bibfield  {author} {\bibinfo {author} {\bibfnamefont {I.}~\bibnamefont
  {Su\'arez~Fern\'andez}}, \bibinfo {author} {\bibfnamefont {R.}~\bibnamefont
  {Vicente}},\ and\ \bibinfo {author} {\bibfnamefont {D.}~\bibnamefont
  {Hilditch}},\ }\bibfield  {title} {\bibinfo {title} {Semilinear wave model
  for critical collapse},\ }\href {https://doi.org/10.1103/PhysRevD.103.044016}
  {\bibfield  {journal} {\bibinfo  {journal} {Phys. Rev. D}\ }\textbf {\bibinfo
  {volume} {103}},\ \bibinfo {pages} {044016} (\bibinfo {year}
  {2021}{\natexlab{a}})}\BibitemShut {NoStop}%
\bibitem [{\citenamefont {Brill}(1959)}]{Bri59}%
  \BibitemOpen
  \bibfield  {author} {\bibinfo {author} {\bibfnamefont {D.~S.}\ \bibnamefont
  {Brill}},\ }\bibfield  {title} {\bibinfo {title} {On the positive definite
  mass of the {B}ondi-{W}eber-{W}heeler time-symmetric gravitational waves},\
  }\href@noop {} {\bibfield  {journal} {\bibinfo  {journal} {Ann. Phys. (N.
  Y.)}\ }\textbf {\bibinfo {volume} {7}},\ \bibinfo {pages} {466} (\bibinfo
  {year} {1959})}\BibitemShut {NoStop}%
\bibitem [{\citenamefont {Teukolsky}(1982)}]{Teu82}%
  \BibitemOpen
  \bibfield  {author} {\bibinfo {author} {\bibfnamefont {S.~A.}\ \bibnamefont
  {Teukolsky}},\ }\bibfield  {title} {\bibinfo {title} {Linearized quadrupole
  waves in general relativity and the motion of test particles},\ }\href@noop
  {} {\bibfield  {journal} {\bibinfo  {journal} {Phys. Rev. D}\ }\textbf
  {\bibinfo {volume} {26}},\ \bibinfo {pages} {745} (\bibinfo {year}
  {1982})}\BibitemShut {NoStop}%
\bibitem [{\citenamefont {{Rinne}}(2008{\natexlab{a}})}]{Rin08b}%
  \BibitemOpen
  \bibfield  {author} {\bibinfo {author} {\bibfnamefont {O.}~\bibnamefont
  {{Rinne}}},\ }\bibfield  {title} {\bibinfo {title} {{Explicit solution of the
  linearized Einstein equations in TT gauge for all multipoles}},\ }\href@noop
  {} {\bibfield  {journal} {\bibinfo  {journal} {ArXiv e-prints}\ } (\bibinfo
  {year} {2008}{\natexlab{a}})},\ \Eprint {https://arxiv.org/abs/0809.1761}
  {arXiv:0809.1761 [gr-qc]} \BibitemShut {NoStop}%
\bibitem [{\citenamefont {Su\'arez~Fern\'andez}\ \emph
  {et~al.}(2021{\natexlab{b}})\citenamefont {Su\'arez~Fern\'andez},
  \citenamefont {Baumgarte},\ and\ \citenamefont {Hilditch}}]{SuaBauHil21}%
  \BibitemOpen
  \bibfield  {author} {\bibinfo {author} {\bibfnamefont {I.}~\bibnamefont
  {Su\'arez~Fern\'andez}}, \bibinfo {author} {\bibfnamefont {T.~W.}\
  \bibnamefont {Baumgarte}},\ and\ \bibinfo {author} {\bibfnamefont
  {D.}~\bibnamefont {Hilditch}},\ }\bibfield  {title} {\bibinfo {title}
  {Comparison of linear brill and teukolsky waves},\ }\href
  {https://doi.org/10.1103/PhysRevD.104.124036} {\bibfield  {journal} {\bibinfo
   {journal} {Phys. Rev. D}\ }\textbf {\bibinfo {volume} {104}},\ \bibinfo
  {pages} {124036} (\bibinfo {year} {2021}{\natexlab{b}})}\BibitemShut
  {NoStop}%
\bibitem [{\citenamefont {Abrahams}\ and\ \citenamefont
  {Evans}(1993)}]{AbrEva93}%
  \BibitemOpen
  \bibfield  {author} {\bibinfo {author} {\bibfnamefont {A.~M.}\ \bibnamefont
  {Abrahams}}\ and\ \bibinfo {author} {\bibfnamefont {C.~R.}\ \bibnamefont
  {Evans}},\ }\bibfield  {title} {\bibinfo {title} {Critical behavior and
  scaling in vacuum axisymmetric gravitational collapse},\ }\href@noop {}
  {\bibfield  {journal} {\bibinfo  {journal} {Phys. Rev. Lett.}\ }\textbf
  {\bibinfo {volume} {70}},\ \bibinfo {pages} {2980} (\bibinfo {year}
  {1993})}\BibitemShut {NoStop}%
\bibitem [{\citenamefont {Abrahams}\ and\ \citenamefont
  {Evans}(1994)}]{AbrEva94}%
  \BibitemOpen
  \bibfield  {author} {\bibinfo {author} {\bibfnamefont {A.~M.}\ \bibnamefont
  {Abrahams}}\ and\ \bibinfo {author} {\bibfnamefont {C.~R.}\ \bibnamefont
  {Evans}},\ }\bibfield  {title} {\bibinfo {title} {{Universality in
  axisymmetric vacuum collapse}},\ }\href
  {https://doi.org/10.1103/PhysRevD.49.3998} {\bibfield  {journal} {\bibinfo
  {journal} {Phys. Rev.}\ }\textbf {\bibinfo {volume} {D49}},\ \bibinfo {pages}
  {3998} (\bibinfo {year} {1994})}\BibitemShut {NoStop}%
%%CITATION = PHRVA,D49,3998;%%
\bibitem [{\citenamefont {Shibata}\ and\ \citenamefont
  {Nakamura}(1995)}]{ShiNak95}%
  \BibitemOpen
  \bibfield  {author} {\bibinfo {author} {\bibfnamefont {M.}~\bibnamefont
  {Shibata}}\ and\ \bibinfo {author} {\bibfnamefont {T.}~\bibnamefont
  {Nakamura}},\ }\bibfield  {title} {\bibinfo {title} {Evolution of
  three-dimensional gravitational waves: {H}armonic slicing case},\ }\href@noop
  {} {\bibfield  {journal} {\bibinfo  {journal} {Phys. Rev. D}\ }\textbf
  {\bibinfo {volume} {52}},\ \bibinfo {pages} {5428} (\bibinfo {year}
  {1995})}\BibitemShut {NoStop}%
\bibitem [{\citenamefont {Bonazzola}\ \emph {et~al.}(2004)\citenamefont
  {Bonazzola}, \citenamefont {Gourgoulhon}, \citenamefont {Grandcl{\'e}ment},\
  and\ \citenamefont {Novak}}]{BonGouGra03}%
  \BibitemOpen
  \bibfield  {author} {\bibinfo {author} {\bibfnamefont {S.}~\bibnamefont
  {Bonazzola}}, \bibinfo {author} {\bibfnamefont {E.}~\bibnamefont
  {Gourgoulhon}}, \bibinfo {author} {\bibfnamefont {P.}~\bibnamefont
  {Grandcl{\'e}ment}},\ and\ \bibinfo {author} {\bibfnamefont {J.}~\bibnamefont
  {Novak}},\ }\bibfield  {title} {\bibinfo {title} {A constrained scheme for
  {E}instein equations based on {D}irac gauge and spherical coordinates},\
  }\href {https://doi.org/10.1103/PhysRevD.70.104007} {\bibfield  {journal}
  {\bibinfo  {journal} {Phys. Rev. D}\ }\textbf {\bibinfo {volume} {70}},\
  \bibinfo {pages} {104007} (\bibinfo {year} {2004})},\ \Eprint
  {https://arxiv.org/abs/gr-qc/0307082} {arXiv:gr-qc/0307082} \BibitemShut
  {NoStop}%
%%CITATION = GR-QC 0307082;%%
\bibitem [{\citenamefont {Pfeiffer}\ \emph {et~al.}(2005)\citenamefont
  {Pfeiffer}, \citenamefont {Kidder}, \citenamefont {Scheel},\ and\
  \citenamefont {Shoemaker}}]{PfeKidSch04}%
  \BibitemOpen
  \bibfield  {author} {\bibinfo {author} {\bibfnamefont {H.~P.}\ \bibnamefont
  {Pfeiffer}}, \bibinfo {author} {\bibfnamefont {L.~E.}\ \bibnamefont
  {Kidder}}, \bibinfo {author} {\bibfnamefont {M.~A.}\ \bibnamefont {Scheel}},\
  and\ \bibinfo {author} {\bibfnamefont {D.}~\bibnamefont {Shoemaker}},\
  }\bibfield  {title} {\bibinfo {title} {Initial data for {E}instein's
  equations with superposed gravitational waves},\ }\href@noop {} {\bibfield
  {journal} {\bibinfo  {journal} {Phys. Rev. D}\ }\textbf {\bibinfo {volume}
  {71}},\ \bibinfo {pages} {024020} (\bibinfo {year} {2005})},\ \Eprint
  {https://arxiv.org/abs/gr-qc/0410016} {gr-qc/0410016} \BibitemShut {NoStop}%
%%CITATION = GR-QC 0410016;%%
\bibitem [{\citenamefont {Hilditch}\ \emph {et~al.}(2013)\citenamefont
  {Hilditch}, \citenamefont {Baumgarte}, \citenamefont {Weyhausen},
  \citenamefont {Dietrich}, \citenamefont {Br{\"u}gmann}, \citenamefont
  {Montero},\ and\ \citenamefont {M{\"u}ller}}]{HilBauWey13}%
  \BibitemOpen
  \bibfield  {author} {\bibinfo {author} {\bibfnamefont {D.}~\bibnamefont
  {Hilditch}}, \bibinfo {author} {\bibfnamefont {T.~W.}\ \bibnamefont
  {Baumgarte}}, \bibinfo {author} {\bibfnamefont {A.}~\bibnamefont
  {Weyhausen}}, \bibinfo {author} {\bibfnamefont {T.}~\bibnamefont {Dietrich}},
  \bibinfo {author} {\bibfnamefont {B.}~\bibnamefont {Br{\"u}gmann}}, \bibinfo
  {author} {\bibfnamefont {P.~J.}\ \bibnamefont {Montero}},\ and\ \bibinfo
  {author} {\bibfnamefont {E.}~\bibnamefont {M{\"u}ller}},\ }\bibfield  {title}
  {\bibinfo {title} {Collapse of nonlinear gravitational waves in
  moving-puncture coordinates},\ }\href@noop {} {\bibfield  {journal} {\bibinfo
   {journal} {Phys.Rev.}\ }\textbf {\bibinfo {volume} {D88}},\ \bibinfo {pages}
  {103009} (\bibinfo {year} {2013})},\ \Eprint
  {https://arxiv.org/abs/1309.5008} {arXiv:1309.5008 [gr-qc]} \BibitemShut
  {NoStop}%
%%CITATION = ARXIV:1309.5008;%%
\bibitem [{\citenamefont {Alcubierre}\ \emph {et~al.}(2000)\citenamefont
  {Alcubierre}, \citenamefont {Allen}, \citenamefont {Br{\"u}gmann},
  \citenamefont {Lanfermann}, \citenamefont {Seidel}, \citenamefont {Suen},\
  and\ \citenamefont {Tobias}}]{AlcAllBru99a}%
  \BibitemOpen
  \bibfield  {author} {\bibinfo {author} {\bibfnamefont {M.}~\bibnamefont
  {Alcubierre}}, \bibinfo {author} {\bibfnamefont {G.}~\bibnamefont {Allen}},
  \bibinfo {author} {\bibfnamefont {B.}~\bibnamefont {Br{\"u}gmann}}, \bibinfo
  {author} {\bibfnamefont {G.}~\bibnamefont {Lanfermann}}, \bibinfo {author}
  {\bibfnamefont {E.}~\bibnamefont {Seidel}}, \bibinfo {author} {\bibfnamefont
  {W.-M.}\ \bibnamefont {Suen}},\ and\ \bibinfo {author} {\bibfnamefont
  {M.}~\bibnamefont {Tobias}},\ }\bibfield  {title} {\bibinfo {title}
  {Gravitational collapse of gravitational waves in {3D} numerical
  relativity},\ }\href@noop {} {\bibfield  {journal} {\bibinfo  {journal}
  {Phys. Rev. D}\ }\textbf {\bibinfo {volume} {61}},\ \bibinfo {pages} {041501
  (R)} (\bibinfo {year} {2000})},\ \Eprint
  {https://arxiv.org/abs/gr-qc/9904013} {gr-qc/9904013} \BibitemShut {NoStop}%
\bibitem [{\citenamefont {Garfinkle}\ and\ \citenamefont
  {Duncan}(2001)}]{GarDun00}%
  \BibitemOpen
  \bibfield  {author} {\bibinfo {author} {\bibfnamefont {D.}~\bibnamefont
  {Garfinkle}}\ and\ \bibinfo {author} {\bibfnamefont {G.~C.}\ \bibnamefont
  {Duncan}},\ }\bibfield  {title} {\bibinfo {title} {Numerical evolution of
  {B}rill waves},\ }\href@noop {} {\bibfield  {journal} {\bibinfo  {journal}
  {Phys. Rev. D}\ }\textbf {\bibinfo {volume} {63}},\ \bibinfo {pages} {044011}
  (\bibinfo {year} {2001})},\ \Eprint {https://arxiv.org/abs/gr-qc/0006073}
  {gr-qc/0006073} \BibitemShut {NoStop}%
%%CITATION = PHRVA,D63,044011;%%
\bibitem [{\citenamefont {Santamaria}(2006)}]{San06}%
  \BibitemOpen
  \bibfield  {author} {\bibinfo {author} {\bibfnamefont {L.}~\bibnamefont
  {Santamaria}},\ }\emph {\bibinfo {title} {Nonlinear 3D Evolutions of
  {B}rillwave Sapcetimes and Critical Phenomena}},\ \href@noop {} {Master's
  thesis},\ \bibinfo  {school} {Friedrich-Schiller-Universit{\"a}t Jena}
  (\bibinfo {year} {2006})\BibitemShut {NoStop}%
\bibitem [{\citenamefont {{Rinne}}(2008{\natexlab{b}})}]{Rin08}%
  \BibitemOpen
  \bibfield  {author} {\bibinfo {author} {\bibfnamefont {O.}~\bibnamefont
  {{Rinne}}},\ }\bibfield  {title} {\bibinfo {title} {{Constrained evolution in
  axisymmetry and the gravitational collapse of prolate Brill waves}},\ }\href
  {https://doi.org/10.1088/0264-9381/25/13/135009} {\bibfield  {journal}
  {\bibinfo  {journal} {Classical and Quantum Gravity}\ }\textbf {\bibinfo
  {volume} {25}},\ \bibinfo {eid} {135009} (\bibinfo {year}
  {2008}{\natexlab{b}})},\ \Eprint {https://arxiv.org/abs/0802.3791}
  {arXiv:0802.3791 [gr-qc]} \BibitemShut {NoStop}%
\bibitem [{\citenamefont {Sorkin}(2011)}]{Sor10}%
  \BibitemOpen
  \bibfield  {author} {\bibinfo {author} {\bibfnamefont {E.}~\bibnamefont
  {Sorkin}},\ }\bibfield  {title} {\bibinfo {title} {{On critical collapse of
  gravitational waves}},\ }\href
  {https://doi.org/10.1088/0264-9381/28/2/025011} {\bibfield  {journal}
  {\bibinfo  {journal} {Class. Quant. Grav.}\ }\textbf {\bibinfo {volume}
  {28}},\ \bibinfo {pages} {025011} (\bibinfo {year} {2011})},\ \Eprint
  {https://arxiv.org/abs/1008.3319} {arXiv:1008.3319 [gr-qc]} \BibitemShut
  {NoStop}%
%%CITATION = ARXIV:1008.3319;%%
\bibitem [{\citenamefont {Hilditch}\ \emph {et~al.}(2017)\citenamefont
  {Hilditch}, \citenamefont {Weyhausen},\ and\ \citenamefont
  {Br{\"u}gmann}}]{HilWeyBru17}%
  \BibitemOpen
  \bibfield  {author} {\bibinfo {author} {\bibfnamefont {D.}~\bibnamefont
  {Hilditch}}, \bibinfo {author} {\bibfnamefont {A.}~\bibnamefont
  {Weyhausen}},\ and\ \bibinfo {author} {\bibfnamefont {B.}~\bibnamefont
  {Br{\"u}gmann}},\ }\bibfield  {title} {\bibinfo {title} {{Evolutions of
  centered Brill waves with a pseudospectral method}},\ }\href@noop {}
  {\bibfield  {journal} {\bibinfo  {journal} {Phys. Rev.}\ }\textbf {\bibinfo
  {volume} {D96}},\ \bibinfo {pages} {104051} (\bibinfo {year} {2017})},\
  \Eprint {https://arxiv.org/abs/1706.01829} {arXiv:1706.01829 [gr-qc]}
  \BibitemShut {NoStop}%
%%CITATION = ARXIV:1706.01829;%%
\bibitem [{\citenamefont {Khirnov}\ and\ \citenamefont
  {Ledvinka}(2018)}]{KhiLed18}%
  \BibitemOpen
  \bibfield  {author} {\bibinfo {author} {\bibfnamefont {A.}~\bibnamefont
  {Khirnov}}\ and\ \bibinfo {author} {\bibfnamefont {T.}~\bibnamefont
  {Ledvinka}},\ }\bibfield  {title} {\bibinfo {title} {{Slicing conditions for
  axisymmetric gravitational collapse of Brill waves}},\ }\href@noop {}
  {\bibfield  {journal} {\bibinfo  {journal} {Class. Quant. Grav.}\ }\textbf
  {\bibinfo {volume} {35}},\ \bibinfo {pages} {215003} (\bibinfo {year}
  {2018})}\BibitemShut {NoStop}%
%%CITATION = CQGRD,35,215003;%%
\bibitem [{\citenamefont {Ledvinka}\ and\ \citenamefont
  {Khirnov}(2021)}]{LedKhi21}%
  \BibitemOpen
  \bibfield  {author} {\bibinfo {author} {\bibfnamefont {T.}~\bibnamefont
  {Ledvinka}}\ and\ \bibinfo {author} {\bibfnamefont {A.}~\bibnamefont
  {Khirnov}},\ }\bibfield  {title} {\bibinfo {title} {Universality of curvature
  invariants in critical vacuum gravitational collapse},\ }\href
  {https://doi.org/10.1103/PhysRevLett.127.011104} {\bibfield  {journal}
  {\bibinfo  {journal} {Phys. Rev. Lett.}\ }\textbf {\bibinfo {volume} {127}},\
  \bibinfo {pages} {011104} (\bibinfo {year} {2021})}\BibitemShut {NoStop}%
\bibitem [{\citenamefont {Br{\"u}gmann}(2013)}]{Bru11}%
  \BibitemOpen
  \bibfield  {author} {\bibinfo {author} {\bibfnamefont {B.}~\bibnamefont
  {Br{\"u}gmann}},\ }\bibfield  {title} {\bibinfo {title} {A pseudospectral
  matrix method for time-dependent tensor fields on a spherical shell},\
  }\href@noop {} {\bibfield  {journal} {\bibinfo  {journal} {J. Comput. Phys.}\
  }\textbf {\bibinfo {volume} {235}},\ \bibinfo {pages} {216} (\bibinfo {year}
  {2013})},\ \Eprint {https://arxiv.org/abs/1104.3408} {arXiv:1104.3408
  [physics.comp-ph]} \BibitemShut {NoStop}%
%%CITATION = ARXIV:1104.3408;%%
\bibitem [{\citenamefont {Hilditch}\ \emph {et~al.}(2016)\citenamefont
  {Hilditch}, \citenamefont {Weyhausen},\ and\ \citenamefont
  {Br{\"u}gmann}}]{HilWeyBru15}%
  \BibitemOpen
  \bibfield  {author} {\bibinfo {author} {\bibfnamefont {D.}~\bibnamefont
  {Hilditch}}, \bibinfo {author} {\bibfnamefont {A.}~\bibnamefont
  {Weyhausen}},\ and\ \bibinfo {author} {\bibfnamefont {B.}~\bibnamefont
  {Br{\"u}gmann}},\ }\bibfield  {title} {\bibinfo {title} {{Pseudospectral
  method for gravitational wave collapse}},\ }\href
  {https://doi.org/10.1103/PhysRevD.93.063006} {\bibfield  {journal} {\bibinfo
  {journal} {Phys. Rev.}\ }\textbf {\bibinfo {volume} {D93}},\ \bibinfo {pages}
  {063006} (\bibinfo {year} {2016})},\ \Eprint
  {https://arxiv.org/abs/1504.04732} {arXiv:1504.04732 [gr-qc]} \BibitemShut
  {NoStop}%
%%CITATION = ARXIV:1504.04732;%%
\bibitem [{\citenamefont {Lindblom}\ \emph {et~al.}(2006)\citenamefont
  {Lindblom}, \citenamefont {Scheel}, \citenamefont {Kidder}, \citenamefont
  {Owen},\ and\ \citenamefont {Rinne}}]{LinSchKid05}%
  \BibitemOpen
  \bibfield  {author} {\bibinfo {author} {\bibfnamefont {L.}~\bibnamefont
  {Lindblom}}, \bibinfo {author} {\bibfnamefont {M.~A.}\ \bibnamefont
  {Scheel}}, \bibinfo {author} {\bibfnamefont {L.~E.}\ \bibnamefont {Kidder}},
  \bibinfo {author} {\bibfnamefont {R.}~\bibnamefont {Owen}},\ and\ \bibinfo
  {author} {\bibfnamefont {O.}~\bibnamefont {Rinne}},\ }\bibfield  {title}
  {\bibinfo {title} {A new generalized harmonic evolution system},\ }\href@noop
  {} {\bibfield  {journal} {\bibinfo  {journal} {Class. Quant. Grav.}\ }\textbf
  {\bibinfo {volume} {23}},\ \bibinfo {pages} {S447} (\bibinfo {year}
  {2006})},\ \Eprint {https://arxiv.org/abs/gr-qc/0512093} {gr-qc/0512093}
  \BibitemShut {NoStop}%
%%CITATION = GR-QC 0512093;%%
\bibitem [{\citenamefont {Rinne}(2006)}]{Rin06a}%
  \BibitemOpen
  \bibfield  {author} {\bibinfo {author} {\bibfnamefont {O.}~\bibnamefont
  {Rinne}},\ }\bibfield  {title} {\bibinfo {title} {{Stable
  radiation-controlling boundary conditions for the generalized harmonic
  Einstein equations}},\ }\href {https://doi.org/10.1088/0264-9381/23/22/013}
  {\bibfield  {journal} {\bibinfo  {journal} {Class. Quant. Grav.}\ }\textbf
  {\bibinfo {volume} {23}},\ \bibinfo {pages} {6275} (\bibinfo {year}
  {2006})},\ \Eprint {https://arxiv.org/abs/gr-qc/0606053}
  {arXiv:gr-qc/0606053} \BibitemShut {NoStop}%
%%CITATION = GR-QC/0606053;%%
\bibitem [{\citenamefont {Alcubierre}\ \emph {et~al.}(2001)\citenamefont
  {Alcubierre}, \citenamefont {Brandt}, \citenamefont {Br{\"u}gmann},
  \citenamefont {Holz}, \citenamefont {Seidel}, \citenamefont {Takahashi},\
  and\ \citenamefont {Thornburg}}]{AlcBraBru99a}%
  \BibitemOpen
  \bibfield  {author} {\bibinfo {author} {\bibfnamefont {M.}~\bibnamefont
  {Alcubierre}}, \bibinfo {author} {\bibfnamefont {S.~R.}\ \bibnamefont
  {Brandt}}, \bibinfo {author} {\bibfnamefont {B.}~\bibnamefont
  {Br{\"u}gmann}}, \bibinfo {author} {\bibfnamefont {D.}~\bibnamefont {Holz}},
  \bibinfo {author} {\bibfnamefont {E.}~\bibnamefont {Seidel}}, \bibinfo
  {author} {\bibfnamefont {R.}~\bibnamefont {Takahashi}},\ and\ \bibinfo
  {author} {\bibfnamefont {J.}~\bibnamefont {Thornburg}},\ }\bibfield  {title}
  {\bibinfo {title} {Symmetry without symmetry: Numerical simulation of
  axisymmetric systems using {C}artesian grids},\ }\href@noop {} {\bibfield
  {journal} {\bibinfo  {journal} {Int. J. Mod. Phys. D}\ }\textbf {\bibinfo
  {volume} {10}},\ \bibinfo {pages} {273} (\bibinfo {year} {2001})},\ \Eprint
  {https://arxiv.org/abs/gr-qc/9908012} {gr-qc/9908012} \BibitemShut {NoStop}%
\bibitem [{\citenamefont {Pretorius}(2005)}]{Pre05}%
  \BibitemOpen
  \bibfield  {author} {\bibinfo {author} {\bibfnamefont {F.}~\bibnamefont
  {Pretorius}},\ }\bibfield  {title} {\bibinfo {title} {Evolution of binary
  black hole spacetimes},\ }\href@noop {} {\bibfield  {journal} {\bibinfo
  {journal} {Phys. Rev. Lett.}\ }\textbf {\bibinfo {volume} {95}},\ \bibinfo
  {pages} {121101} (\bibinfo {year} {2005})},\ \Eprint
  {https://arxiv.org/abs/gr-qc/0507014} {gr-qc/0507014} \BibitemShut {NoStop}%
%%CITATION = GR-QC 0507014;%%
\bibitem [{\citenamefont {Hesthaven}(2000)}]{Hes00}%
  \BibitemOpen
  \bibfield  {author} {\bibinfo {author} {\bibfnamefont {J.~S.}\ \bibnamefont
  {Hesthaven}},\ }\bibfield  {title} {\bibinfo {title} {Spectral penalty
  methods},\ }\href@noop {} {\bibfield  {journal} {\bibinfo  {journal} {Appl.
  Numer. Math.}\ }\textbf {\bibinfo {volume} {33}},\ \bibinfo {pages} {23}
  (\bibinfo {year} {2000})}\BibitemShut {NoStop}%
\bibitem [{\citenamefont {Hesthaven}\ \emph {et~al.}(2007)\citenamefont
  {Hesthaven}, \citenamefont {Gottlieb},\ and\ \citenamefont
  {Gottlieb}}]{HesGotGot07}%
  \BibitemOpen
  \bibfield  {author} {\bibinfo {author} {\bibfnamefont {J.~S.}\ \bibnamefont
  {Hesthaven}}, \bibinfo {author} {\bibfnamefont {S.}~\bibnamefont
  {Gottlieb}},\ and\ \bibinfo {author} {\bibfnamefont {D.}~\bibnamefont
  {Gottlieb}},\ }\href@noop {} {\emph {\bibinfo {title} {Spectral Methods for
  Time-Dependent Problems}}}\ (\bibinfo  {publisher} {Cambridge University
  Press},\ \bibinfo {address} {Cambridge},\ \bibinfo {year} {2007})\BibitemShut
  {NoStop}%
\bibitem [{\citenamefont {Taylor}\ \emph {et~al.}(2010)\citenamefont {Taylor},
  \citenamefont {Kidder},\ and\ \citenamefont {Teukolsky}}]{TayKidTeu10}%
  \BibitemOpen
  \bibfield  {author} {\bibinfo {author} {\bibfnamefont {N.~W.}\ \bibnamefont
  {Taylor}}, \bibinfo {author} {\bibfnamefont {L.~E.}\ \bibnamefont {Kidder}},\
  and\ \bibinfo {author} {\bibfnamefont {S.~A.}\ \bibnamefont {Teukolsky}},\
  }\bibfield  {title} {\bibinfo {title} {{Spectral methods for the wave
  equation in second-order form}},\ }\href
  {https://doi.org/10.1103/PhysRevD.82.024037} {\bibfield  {journal} {\bibinfo
  {journal} {Phys.Rev.}\ }\textbf {\bibinfo {volume} {D82}},\ \bibinfo {pages}
  {024037} (\bibinfo {year} {2010})},\ \Eprint
  {https://arxiv.org/abs/1005.2922} {arXiv:1005.2922 [gr-qc]} \BibitemShut
  {NoStop}%
%%CITATION = ARXIV:1005.2922;%%
\bibitem [{\citenamefont {Holz}\ \emph {et~al.}(1993)\citenamefont {Holz},
  \citenamefont {Miller}, \citenamefont {Wakano},\ and\ \citenamefont
  {Wheeler}}]{HolMilWak93}%
  \BibitemOpen
  \bibfield  {author} {\bibinfo {author} {\bibfnamefont {D.}~\bibnamefont
  {Holz}}, \bibinfo {author} {\bibfnamefont {W.}~\bibnamefont {Miller}},
  \bibinfo {author} {\bibfnamefont {M.}~\bibnamefont {Wakano}},\ and\ \bibinfo
  {author} {\bibfnamefont {J.}~\bibnamefont {Wheeler}},\ }in\ \href@noop {}
  {\emph {\bibinfo {booktitle} {Directions in General Relativity: Proceedings
  of the 1993 International Symposium, {M}aryland; Papers in honor of Dieter
  {B}rill}}},\ \bibinfo {editor} {edited by\ \bibinfo {editor} {\bibfnamefont
  {B.}~\bibnamefont {Hu}}\ and\ \bibinfo {editor} {\bibfnamefont
  {T.}~\bibnamefont {Jacobson}}}\ (\bibinfo  {publisher} {Cambridge University
  Press},\ \bibinfo {address} {Cambridge, England},\ \bibinfo {year}
  {1993})\BibitemShut {NoStop}%
\bibitem [{\citenamefont {Renkhoff}()}]{ahloc3d}%
  \BibitemOpen
  \bibfield  {author} {\bibinfo {author} {\bibfnamefont {S.}~\bibnamefont
  {Renkhoff}},\ }\href@noop {} {}\bibinfo {note}
  {\url{https://git.tpi.uni-jena.de/srenkhoff/ahloc3d}}\BibitemShut {NoStop}%
\bibitem [{\citenamefont {Schnetter}(2002)}]{Sch02}%
  \BibitemOpen
  \bibfield  {author} {\bibinfo {author} {\bibfnamefont {E.}~\bibnamefont
  {Schnetter}},\ }\bibfield  {title} {\bibinfo {title} {A fast apparent horizon
  algorithm},\ }\Eprint {https://arxiv.org/abs/gr-qc/0206003} {gr-qc/0206003}
  (\bibinfo {year} {2002}),\ \bibinfo {note} {gr-qc/0206003}\BibitemShut
  {NoStop}%
\bibitem [{\citenamefont {Tod}(1991)}]{Tod91}%
  \BibitemOpen
  \bibfield  {author} {\bibinfo {author} {\bibfnamefont {K.~P.}\ \bibnamefont
  {Tod}},\ }\bibfield  {title} {\bibinfo {title} {Looking for marginally
  trapped surfaces},\ }\href@noop {} {\bibfield  {journal} {\bibinfo  {journal}
  {Class. Quantum Grav.}\ }\textbf {\bibinfo {volume} {8}},\ \bibinfo {pages}
  {L115} (\bibinfo {year} {1991})}\BibitemShut {NoStop}%
\bibitem [{\citenamefont {Khirnov}(2021)}]{Khi21}%
  \BibitemOpen
  \bibfield  {author} {\bibinfo {author} {\bibfnamefont {A.}~\bibnamefont
  {Khirnov}},\ }\emph {\bibinfo {title} {Representation of dynamical black hole
  spacetimes in numerical simulations}},\ \href@noop {} {Ph.D. thesis},\
  \bibinfo  {school} {Charles University, Prague} (\bibinfo {year}
  {2021})\BibitemShut {NoStop}%
\bibitem [{\citenamefont {Hilditch}(2015)}]{Hil15}%
  \BibitemOpen
  \bibfield  {author} {\bibinfo {author} {\bibfnamefont {D.}~\bibnamefont
  {Hilditch}},\ }\bibfield  {title} {\bibinfo {title} {{Dual Foliation
  Formulations of General Relativity}},\ }\href@noop {} {\  (\bibinfo {year}
  {2015})},\ \Eprint {https://arxiv.org/abs/1509.02071} {arXiv:1509.02071
  [gr-qc]} \BibitemShut {NoStop}%
%%CITATION = ARXIV:1509.02071;%%
\bibitem [{\citenamefont {Hilditch}\ \emph {et~al.}(2018)\citenamefont
  {Hilditch}, \citenamefont {Harms}, \citenamefont {Bugner}, \citenamefont
  {R{\"u}ter},\ and\ \citenamefont {Br{\"u}gmann}}]{HilHarBug16}%
  \BibitemOpen
  \bibfield  {author} {\bibinfo {author} {\bibfnamefont {D.}~\bibnamefont
  {Hilditch}}, \bibinfo {author} {\bibfnamefont {E.}~\bibnamefont {Harms}},
  \bibinfo {author} {\bibfnamefont {M.}~\bibnamefont {Bugner}}, \bibinfo
  {author} {\bibfnamefont {H.}~\bibnamefont {R{\"u}ter}},\ and\ \bibinfo
  {author} {\bibfnamefont {B.}~\bibnamefont {Br{\"u}gmann}},\ }\bibfield
  {title} {\bibinfo {title} {{The evolution of hyperboloidal data with the dual
  foliation formalism: Mathematical analysis and wave equation tests}},\ }\href
  {https://doi.org/10.1088/1361-6382/aaa4ac} {\bibfield  {journal} {\bibinfo
  {journal} {Class. Quant. Grav.}\ }\textbf {\bibinfo {volume} {35}},\ \bibinfo
  {pages} {055003} (\bibinfo {year} {2018})},\ \Eprint
  {https://arxiv.org/abs/1609.08949} {arXiv:1609.08949 [gr-qc]} \BibitemShut
  {NoStop}%
%%CITATION = ARXIV:1609.08949;%%
\bibitem [{\citenamefont {Bhattacharyya}\ \emph {et~al.}(2021)\citenamefont
  {Bhattacharyya}, \citenamefont {Hilditch}, \citenamefont {Rajesh~Nayak},
  \citenamefont {Renkhoff}, \citenamefont {R\"uter},\ and\ \citenamefont
  {Br\"ugmann}}]{BhaHilRaj21}%
  \BibitemOpen
  \bibfield  {author} {\bibinfo {author} {\bibfnamefont {M.~K.}\ \bibnamefont
  {Bhattacharyya}}, \bibinfo {author} {\bibfnamefont {D.}~\bibnamefont
  {Hilditch}}, \bibinfo {author} {\bibfnamefont {K.}~\bibnamefont
  {Rajesh~Nayak}}, \bibinfo {author} {\bibfnamefont {S.}~\bibnamefont
  {Renkhoff}}, \bibinfo {author} {\bibfnamefont {H.~R.}\ \bibnamefont
  {R\"uter}},\ and\ \bibinfo {author} {\bibfnamefont {B.}~\bibnamefont
  {Br\"ugmann}},\ }\bibfield  {title} {\bibinfo {title} {{Implementation of the
  dual foliation generalized harmonic gauge formulation with application to
  spherical black hole excision}},\ }\href
  {https://doi.org/10.1103/PhysRevD.103.064072} {\bibfield  {journal} {\bibinfo
   {journal} {Phys. Rev. D}\ }\textbf {\bibinfo {volume} {103}},\ \bibinfo
  {pages} {064072} (\bibinfo {year} {2021})},\ \Eprint
  {https://arxiv.org/abs/2101.12094} {arXiv:2101.12094 [gr-qc]} \BibitemShut
  {NoStop}%
\bibitem [{\citenamefont {Garfinkle}\ and\ \citenamefont
  {Pretorius}(2020)}]{GarPre20}%
  \BibitemOpen
  \bibfield  {author} {\bibinfo {author} {\bibfnamefont {D.}~\bibnamefont
  {Garfinkle}}\ and\ \bibinfo {author} {\bibfnamefont {F.}~\bibnamefont
  {Pretorius}},\ }\bibfield  {title} {\bibinfo {title} {Spike behavior in the
  approach to spacetime singularities},\ }\href
  {https://doi.org/10.1103/PhysRevD.102.124067} {\bibfield  {journal} {\bibinfo
   {journal} {Phys. Rev. D}\ }\textbf {\bibinfo {volume} {102}},\ \bibinfo
  {pages} {124067} (\bibinfo {year} {2020})}\BibitemShut {NoStop}%
\end{thebibliography}%
%=========================================================

%=========================================================
\end{document}